\documentclass[12pt]{article}

\pdfoutput=1

\usepackage[top=80pt,bottom=85pt,left=85pt,right=85pt]{geometry}
\usepackage{amssymb}
\usepackage{amsmath}
\usepackage{setspace}
\usepackage{sectsty}
\usepackage{cite}
\usepackage{bbm}
\usepackage[utf8]{inputenc}
\usepackage[T1]{fontenc}
\usepackage{comment}
\usepackage[english]{babel}

\usepackage[usenames]{xcolor}
\usepackage{graphicx,subcaption}
\usepackage{setspace}
\usepackage{float}
\usepackage{booktabs}
\usepackage[debug,pageanchor=false]{hyperref}
\definecolor{link}{rgb}{.8,.15,.1}
\definecolor{pigment}{rgb}{0.36, 0.54, 0.66}
\definecolor{pigment2}{rgb}{0.19, 0.55, 0.91}
\definecolor{pigment3}{rgb}{0.2, 0.2, 0.6}
\definecolor{light-gray}{gray}{0.75}
\hypersetup{colorlinks=true,linkcolor=link,citecolor=link,urlcolor=link,linktocpage}

\usepackage{array,multirow,booktabs,longtable}
\usepackage{mathrsfs}
\usepackage{tikz-cd} 
\usetikzlibrary{backgrounds, arrows,calc,shapes,decorations.pathreplacing, decorations.markings, automata,positioning}

\tikzset{
        cvertex/.style={circle,draw=black,inner sep=1pt,outer sep=3pt},
        vertex/.style={circle,fill=black,inner sep=1pt,outer sep=3pt},
        star/.style={circle,fill=yellow,inner sep=0.75pt,outer sep=0.75pt},
        tvertex/.style={inner sep=1pt,font=\scriptsize},
        gap/.style={inner sep=0.5pt,fill=white}}

\tikzstyle{mybox} = [draw=black, fill=blue!10, very thick,
    rectangle, rounded corners, inner sep=10pt, inner ysep=20pt]
\tikzstyle{boxtitle} =[fill=blue!50, text=white,rectangle,rounded corners]

\newcommand{\cc}{\mathbb{C}}

\newcommand{\pp}{\mathbb{P}}

\def\cL{\mathcal{L}}
\def\cV{\mathcal{V}}
\def\cO{\mathcal{O}}

\DeclareMathOperator{\SU}{SU}
\DeclareMathOperator{\U}{U}

\DeclareMathOperator{\GL}{GL}
\DeclareMathOperator{\SL}{SL}
\DeclareMathOperator{\im}{im}
\DeclareMathOperator{\coker}{coker}

\DeclareMathOperator{\End}{End}
\DeclareMathOperator{\Hom}{Hom}

\newcommand{\todo}[1]{}
\renewcommand{\todo}[1]{{\color{red} TODO: {#1}}}

\newcommand{\be}{\begin{equation}}  
\newcommand{\ee}{\end{equation}}  
\newcommand{\bea}{\begin{align}}
\newcommand{\eea}{\end{align}}
\newcommand{\bp}{\begin{bmatrix*}[r]}  
\newcommand{\ep}{\end{bmatrix*}}  
\newcommand{\bpp}{\begin{bmatrix}}  
\newcommand{\epp}{\end{bmatrix}}  
\newcommand{\bcd}{\begin{center}
\begin{tikzcd}}
\newcommand{\ecd}{\end{tikzcd} \end{center}}

\makeatletter
\@addtoreset{equation}{section}
\makeatother

\begin{document}


\begin{titlepage}

\begin{center}

\vskip .3in \noindent

{\Large \bf{Geometric engineering on flops of length two}}

\bigskip\bigskip

Andr\'es Collinucci$^a$, Marco Fazzi$^{b,c}$, and Roberto Valandro$^{d,e,f}$ \\

\bigskip


\bigskip
{\footnotesize
 \it

$^a$ Service de Physique Th\'eorique et Math\'ematique, Universit\'e Libre de Bruxelles and \\ International Solvay Institutes, Campus Plaine C.P.~231, B-1050 Bruxelles, Belgium\\
\vspace{.25cm}
$^b$ Department of Physics, Technion, 32000 Haifa, Israel\\
\vspace{.25cm}
$^c$ Department of Mathematics and Haifa Research Center for Theoretical Physics and Astrophysics, University of Haifa, 31905 Haifa, Israel\\
\vspace{.25cm}
$^d$ Dipartimento di Fisica, Universit\`a di Trieste, Strada Costiera 11, I-34151 Trieste, Italy \\
\vspace{.25cm}
$^e$ INFN, Sezione di Trieste, Via Valerio 2, I-34127 Trieste, Italy \\
\vspace{.25cm}
$^f$ The Abdus Salam International Centre for Theoretical Physics, \\ Strada Costiera 11, I-34151 Trieste, Italy
	
}

\vskip .2cm
{\scriptsize \tt collinucci.phys at gmail.com \hspace{.5cm} mfazzi at physics.technion.ac.il \hspace{.5cm} roberto.valandro at ts.infn.it}



\vskip .7cm
     	{\bf Abstract }
\vskip .1in
\end{center}

Type IIA on the conifold is a prototype example for engineering QED with one charged hypermultiplet. The geometry admits a \emph{flop of length one}. In this paper, we study the next generation of geometric engineering on singular geometries, namely \emph{flops of length two} such as Laufer's example, which we affectionately think of as the \emph{conifold~2.0}. Type IIA on the latter geometry gives QED with higher-charge states. In type IIB, even a single D3-probe gives rise to a nonabelian quiver gauge theory. We study this class of geometries explicitly by leveraging their quiver description, showing how to parametrize the exceptional curve, how to see the flop transition, and how to find the noncompact divisors intersecting the curve. With a view towards F-theory applications, we show how these divisors contribute to the enhancement of the Mordell--Weil group of the local elliptic fibration defined by Laufer's example.

\noindent

\vfill
\eject

\end{titlepage}


\tableofcontents

\newpage 
\section{Introduction} 
\label{sec:intro}

The study of D3-branes at threefold singularities is by now a venerable subject, dating back at least to \cite{Douglas:1996sw, Bershadsky:1998mb, Lawrence:wq}. The worldvolume theory can be inferred to be a four-dimensional $\mathcal{N}=1$ quiver gauge theory describing the fractional branes that probe the singularity of a Calabi--Yau threefold, on which one has defined a type IIB background. Powerful mathematics has been developed to extract a quiver with superpotential describing the field theory directly from the singularity. These techniques are greatly simplified in favorable situations, namely when the singularity is an orbifold or toric \cite{Franco:2005rj,Hanany:2005ve,Franco:2005sm}. The quintessential example of the latter class is the conifold $\mathcal{C}$, 
\begin{equation}
P := x^2 - y^2 - z\,t =0\ \subset \ \cc^4\ , \nonumber
\end{equation}
about which a great deal is known both in field theory \cite{klebanov-witten,Klebanov:2000hb} and in singularity theory \cite{Candelas:1989js}.

Singularities in string theory can also be viewed from a fundamentally different perspective, namely the \emph{geometric engineering} paradigm \cite{katz-klemm-vafa,katz-mayr-vafa}: One defines type IIA on the singularity, and considers the effective field theory arising from the supergravity zero modes on this space, supplemented by the light degrees of freedom created by D2-branes wrapping the vanishing cycles. 
The case of the conifold produces a four-dimensional $\mathcal{N}=2$ SQED with one charged hypermultiplet. The $\U(1)$ gauge group comes from the reduction of the type IIA Ramond--Ramond $C_3$ form,\footnote{Although throughout the paper we will work on noncompact threefolds, one always expects that there be a normalizable harmonic two-form on which to reduce $C_3$. Alternatively, one can consider the singularity as a patch of a compact threefold.} and the charged hyper from a D2 and an anti-D2 wrapped on the exceptional sphere.

However, when the singularity does not belong to either class mentioned above, it is in general more difficult to ``read off'' the field theory from it. One has to resort to the description of so-called B-type (topological) D-branes as objects in the bounded derived category of coherent sheaves on the resolution, pioneered in \cite{Douglas:2000gi,Sharpe:1999qz}. In that context, a practical way of extracting the field theory quiver is to study the \emph{noncommutative crepant resolution} (NCCR henceforth) of the singularity, as discovered in physics in \cite{Berenstein:2001jr} and in mathematics in \cite{Bergh:aa}. A key result by Bondal and Orlov \cite{Bondal:aa} states in fact that the resolved geometry and the NCCR have equivalent (bounded) derived categories. In other words, we can define D-branes on the resolved space by defining modules on the NCCR. 

Knowledge of the singular \emph{coordinate ring} $R=\cc[x,y,z,t]/(P)$ together with a set of special $R$-modules, known as \emph{maximal Cohen--Macaulay} modules (MCM henceforth),\footnote{We also require that the singularity be isolated, and that its coordinate ring $R$ be Gorenstein. In physics language, the latter condition guarantees that the resolved space of the singularity $\text{Spec} R$ be a Calabi--Yau threefold, which is necessary to preserve supersymmetry in four dimensions.} is enough to write down a quiver. In the hypersurface case, the task of finding the quiver and its corresponding superpotential is simplified significantly by harnessing a rather abstract mathematical equivalence of categories introduced by \cite{eisenbud}: The category of MCM $R$-modules is equivalent to that of \emph{matrix factorizations} (MF henceforth) of the singular hypersurface. Suppose $P=0$ is the hypersurface singularity on which we define type IIB with a D3-probe; then an MF of it is a pair of square matrices $(\Phi,\Psi)$ of appropriate dimensionality $n$ such that
\begin{equation}
\Phi \cdot \Psi = \Psi \cdot \Phi = P\, \mathbbm{1}_{n\times n}\ . \nonumber
\end{equation}
These matrices can be thought of as maps between $R^{\oplus n}$ and itself. Then each MF defines concretely an MCM $R$-module via $M := \coker \Psi$; in other words we have an exact sequence
\begin{equation}
\begin{tikzcd}
0 \rar & R^{\oplus n} \rar{\Psi} & R^{\oplus n} \rar{} & M \rar &0\ .
\end{tikzcd}\nonumber
\end{equation}
The NCCR can now be constructed rather explicitly: One essentially ``replaces'' the coordinate ring $R$ of the singularity with the noncommutative ring \cite{Bergh:aa,vdb-flops}
\begin{equation}
A=\text{End}_R (R \oplus M_1 \oplus \ldots \oplus M_r) \ , \nonumber 
\end{equation}
where the summands $M_i$ are a specially selected subset of MCM modules over $R$.\footnote{In the conifold, for instance, there are two such possible modules, and only one of them is chosen.} ($A$ can indeed be understood as a noncommutative enhancement of $R=\End_R R$.)

The advantage of the MF description is that, once appriopriate MF's of the hypersurface are known, the task of computing a quiver with relations (i.e. the cyclic derivatives of the superpotential, or \emph{F-terms}) is completely algorithmic. This approach has been put forward in \cite{aspinwall-morrison}, where many examples of singularities (well-known in the mathematical literature) were used as string theory backgrounds and tackled from the NCCR/MF point of view. The authors of that paper focused in particular on so-called three-dimensional simple flops. These are singular algebraic varieties that admit two crepant resolutions which are birationally isomorphic. The conifold, which is moreover toric, is the simplest example of threefold flop, and can be labeled by an integer $\ell=1$ known as \emph{length}. For our purposes it is enough to characterize this integer as half the size of the matrices in an MF of the singularity.\footnote{See \cite{kollar-length,curto-morrison} for a formal definition.} Indeed the conifold hypersurface equation admits a $2\ell \times 2\ell = 2 \times 2$ MF.

When $\ell=2$ the singularity is no longer toric, which renders the extensive techniques pioneered and developed in \cite{Franco:2005rj,Franco:2005sm,Hanany:2005ve} (among many other works) powerless. However, nontoric singularities are quite generic and moreover higher-length flops make up a much richer class of threefolds \cite{reid}. Thus, providing examples of D3-brane probe theories on such singularities appears as a very interesting challenge in its own right. In this paper, we will focus on a simple flop of length $\ell=2$ known as Laufer's example \cite{laufer}. It is defined as the following hypersurface in $\mathbb{C}^4$:
\begin{equation}
x^2 + y^3 + w z^2+w^{2n+1} y= 0\ , \quad  n \in \mathbb{N}_{> 0}\ . \nonumber
\end{equation}
We will study it thoroughly from the D3-probe perspective, i.e. by exploiting the NCCR technique, which will give us a quiver description of the threefold. However, we will also analyze the field theory that arises from the type IIA geometric engineering point of view. The novelty will be that, in contrast to more familiar singularities with exceptional cycles with normal bundle $\mathcal{O}_{\pp^1}(-1) \oplus \mathcal{O}_{\pp^1}(-1)$ (such as the conifold), this class of singularities is characterized by $\mathcal{O}_{\pp^1}(-3) \oplus \mathcal{O}_{\pp^1}(1)$. In field theory, we will see that this translates to having hypers of charge one and also of charge two.\\

The goal of this paper is two-fold: Firstly, it is meant to be an exposition of the powerful tools from NCCR's to study nontoric, non-orbifold singularities. We will see that we can not only derive the appropriate quiver with relations for the aforementioned length-two flop, but we can also recover the resolved threefold as the moduli space of \emph{stable representations} of the quiver. The peculiarity of this type of threefold is that the associated quiver gauge theory is nonabelian even for a single probe brane.

Secondly, this will be a first example where a flop of length two is described continuously. In \cite{aspinwall-morrison} the flop transition could only be seen as a $\mathbb{Z}_2$ symmetry that exchanges two MF's of the singular hypersurface, akin to the case of the conifold which admits two (small) resolutions via two-by-two MF's. However, the question as to how one can see this from the K\"ahler geometry perspective remained unanswered. How should one describe the exceptional $\mathbb{P}^1$ of the resolution? Can we track it all the way to zero volume, and then see how the flopped curve starts to grow in the other K\"ahler cone? We will accomplish this task explicitly.

Finally, by relying on the NCCR/MF techniques we will also study Weil divisors of the singular geometry, by regarding them as flavor nodes in the quiver, in the spirit of \cite{Forcella:2008au,Franco:2013ana}. This is of particular interest in the context of F-theory or geometric engineering in IIA/M-theory, where such divisors induce extra $\U(1)$ gauge symmetries.\\

This paper is organized as follows. In Section~\ref{sec:coni} we introduce the NCCR and MF techniques and apply them to the well-studied conifold singularity, i.e. the simplest example of length-one flop. In Section~\ref{sec:lauf} we introduce our main case-study, i.e. Laufer's length-two flop, whose NCCR we present in Section~\ref{sec:lauf-nccr}. (In appendix \ref{app:mpi} we present yet another example of a length-two flop.) In Section~\ref{sec:nc-div} we study classes of Weil divisors defined on Laufer's singularity, and we also provide an alternative perspective on them leveraging the M/F-theory duality. In Section~\ref{sec:highercharge} we show that, in the geometric engineering context, these geometries admit higher-charge hypers. We briefly present our conclusions in Section~\ref{sec:conc}.


\section{Warm-up: The conifold threefold} 
\label{sec:coni}

To set the stage, we shall first study the well-known conifold singularity by relying on the powerful NCCR techniques, which we will introduce as we go along in the presentation. Given the general familiarity with the conifold example, the use of the NCCR might seem like an overkill. However, it is instructive to revisit this old example in the less familiar NCCR language, as a warm-up for our main case-study, Laufer's geometry.

\subsection{The singularity and its matrix factorization}
\label{sub:coni-mf}

Consider the well-known Calabi--Yau (CY henceforth) threefold $\mathcal{C}$ defined by the following equation:
\begin{equation}\label{eq:conif-sing}
W_\text{conifold}:\ x^2 - y^2 -tz = 0 \ \subset\  \cc^4\ .
\end{equation}
It has a pointlike singularity at the vanishing locus of the ideal $(x,y,z,t)$ of the coordinate ring $R:=\cc[x,y,z,t]/(x^2 - y^2 -tz)$. 
A threefold will admit a small, K\"ahler, crepant resolution provided there is a Weil (but non-Cartier) divisor. In the conifold case there are two independent Weil divisors, given by the (zero locus of the) following ideals:
\begin{equation}\label{eq:nonC-coni}
(x+y,z) \quad \text{and} \quad  (x-y,z) \ .
\end{equation}
Each of them produces, upon blow-up, a nonsingular threefold. We thus obtain two threefolds $\mathcal{X}_\pm$ related by a simple flop $\mathcal{X}_+ \dashrightarrow \mathcal{X}_-$. This means the two nonsingular varieties are birationally isomorphic away from a subvariety. In the case of simple flops, such subvariety is an irreducible, smooth, rational curve, namely the exceptional $\pp^1$ locus of the resolutions.


As explained in the introduction, given a hypersurface with defining equation $P=0$, a \emph{matrix factorization} (MF) of it is a pair of square matrices $(\Phi,\Psi)$ of appropriate dimensionality $n$ such that
\begin{equation}\label{eq:MFgeneral}
  \Phi\cdot\Psi =  \Psi\cdot\Phi = P \, \mathbbm{1}_{n \times n} \ ,
\end{equation}
where $\mathbbm{1}_{n \times n}$ is the $n\times n$ identity matrix. 
The conifold admits two inequivalent, irreducible, nontrivial MF's, namely $(\Phi,\Psi)$ and $(\Psi,\Phi)$ with
\begin{equation}\label{eq:MFconi}
\Phi=\begin{bmatrix} x-y & -z \\ -t & x+y \end{bmatrix}\ , 
\quad \quad
\Psi= \begin{bmatrix} x+y & z \\ t & x-y \end{bmatrix}\ .
\end{equation}
The two pairs are ordered and inequivalent, i.e. they are not related by similarity transformations. The two Weil divisors \eqref{eq:nonC-coni} are related to the two MF's, as we now explain. First, notice that these matrices can be seen as maps from $R^{\oplus 2}$ to $R^{\oplus 2}$. This allows to construct two modules over $R$ as follows \cite{eisenbud}:
\begin{equation}\label{MCMsConifold}
M := \mbox{coker} \left(R^{\oplus 2} \xrightarrow[]{\Psi} R^{\oplus 2}\right)\ , \quad\quad
M^\vee := \mbox{coker} \left(R^{\oplus 2} \xrightarrow[]{\Phi} R^{\oplus 2}\right)\ .
\end{equation}
For the conifold, these are all the nontrivial irreducible MCM modules up to isomorphism \cite{yoshino}. They are rank-one over the whole conifold threefold except at the singular point, where they are rank-two. As we will see, in the resolved space these pull-back to $\mathcal{O}(1)$ and $\mathcal{O}(-1)$, i.e they become line bundles. The associated divisors are given by the locus where a generic section of the bundle vanishes. These loci can be detected already in the singular space. Consider for instance the map $\Psi$. The locus we are looking for is where a given element of $R^{\oplus 2}$ is inside $\im \Psi$.
Given a generic section
\begin{equation}\label{eq:gensec-coni}
s=\begin{pmatrix} s_1 \\ s_2 \end{pmatrix}
\end{equation}
in $R^{\oplus 2}$, the vanishing locus is where $s$ becomes parallel to the generators of (the rank-one) $\im \Psi$ (i.e. the columns of $\Psi$). This happens for
\begin{equation}\label{eq:det-coni}
  \det \begin{bmatrix} s_1 & x+y \\ s_2 & t \end{bmatrix} = 0\ , \quad\quad
  \det \begin{bmatrix} s_1 & z \\ s_2 & x-y \end{bmatrix} = 0\ ,
\end{equation}
or in other words when 
\begin{equation}\label{eq:Phis-coni}
   \Phi \cdot s = 0 \ .
\end{equation}
Hence we have a whole family $|D_+|$ of non-Cartier divisors parametrized by $(s_1,s_2)$. Notice that for the special choice $s_1=0$, we obtain the locus $(x+y,z)$, that is one of the Weil divisors mentioned in \eqref{eq:nonC-coni}. On the other hand, the divisor $(x-y,z)$ belongs to the family $|D_-|$ defined by
\begin{equation}\label{eq:Psis-coni}
   \Psi \cdot s' = 0 \ ,
\end{equation}
associated with the second MCM module ($M^\vee$).
One can check that the union of $|D_+|$ and $|D_-|$ is in the class of a Cartier divisor.

\subsection{Noncommutative crepant resolution and the quiver}
\label{sub:coni-nccr}

We will now put this knowledge of MCM modules over the conifold to use, by constructing the NCCR of the singularity, and subsequently obtaining its (commutative) small resolution as a quiver moduli space.

\begin{figure}[!ht]
\centering
\includegraphics[scale=1.25]{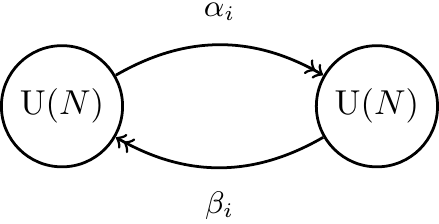}
\caption{Quiver gauge theory of $N$ D3-branes probing the conifold $\mathcal{C}$.}
\label{fig:quiverConifold}
\end{figure}

It is well-known that one can associate a quiver with relations to the conifold singularity. In string theory this can be seen as follows: One considers a stack of $N$ D3-branes at the singular point. The probe theory is described by the quiver in Figure~\ref{fig:quiverConifold} and comes equipped with the (Klebanov--Witten) superpotential \cite{klebanov-witten}:
\begin{equation}\label{eq:kw-superpotential}
\mathcal{W}_\text{KW} = \alpha_1\beta_1\alpha_2\beta_2 - \alpha_1\beta_2\alpha_2\beta_1\ .
\end{equation}
The nodes of the quiver are the fractional brane gauge groups, and the arrows the chiral multiplets charged under such groups.
On the other hand, consider a crepant resolution $\mathcal{X}$ of the conifold (either among $\mathcal{X}_\pm$). Then the bounded derived category of $\mathcal{X}$ is equivalent to the bounded derived category of representations of the quiver described above. One can extract the quiver with superpotential of a given three-dimensional singularity by using Van den Bergh's NCCR's \cite{aspinwall-morrison}. These resolutions require the knowledge of the MCM modules $M_i$ of the singular ring $R$, whose category was shown be equivalent \cite{eisenbud} to a particular category of MF's of the equation defining the singular threefold. Namely, once an MF of the singularity equation $W_\text{conifold}$ is known, we can construct for free its MCM modules via \eqref{MCMsConifold}.

As already mentioned, the basic idea \cite{vdb-nccr} behind the NCCR is to replace the coordinate ring $R = \End_R R$ describing the singular space with the noncommutative ring $A=\End_R (R \oplus M_1 \oplus \ldots \oplus M_r)$ made out of a special subset of the MCM modules. One requires that $A$ be \emph{Cohen--Macaulay}, which is the homological counterpart of crepancy, and that it have finite \emph{global projective dimension}, which essentially means that all projective modules over $A$ admit a finite resolution. That is the homological counterpart of smoothness. (We refer to \cite{wemyss-lec} for a pedagogical introduction to these notions.)

Moreover the ring $A$, which is also an algebra over $R$, can be thought of as the (noncommutative) path algebra of a quiver with relations. Hence, to each summand in $A$ is associated a vertex in the quiver, while the number of arrows from one vertex to another is given by $\dim \Hom_R(M_i, M_j)$ (where $M_i$ can be $R$ too). Once a presentation of the singularity as quiver with relations is known, one can construct a geometric resolution of the singular threefold via the \emph{geometric invariant theory} of King \cite{king-git}.\\

We now apply this procedure to the conifold singularity \eqref{eq:conif-sing} as a first case-study. The NCCR can be characterized by a single MCM module defined by choosing one of the two MF's. Let us take $M=\coker \Psi$ in \eqref{MCMsConifold} for concreteness. The pair
$(\Phi,\Psi)$ is a MF of the singular space, i.e.
\begin{equation}\label{eq:MFlength}
 \Phi\cdot\Psi = \Psi\cdot \Phi = (x^2-y^2-tz) \mathbbm{1}_{2\ell \times 2\ell} \ , 
\end{equation} 
with $\ell=1$. As we have already mentioned elsewhere, we can define the integer $\ell$ to be the length of the flop, which is a numerical invariant that characterizes it \cite{curto-morrison,kollar-length}.

\begin{figure}[!t]
\centering
\includegraphics[scale=1.25]{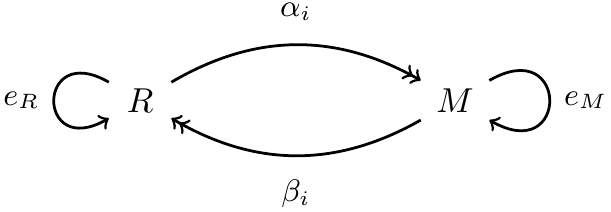}
\caption{The conifold quiver. $e_i$ is the idempotent of node $i=R,M$ in the path algebra of the quiver.}
\label{fig:quiverConifoldNCCR}
\end{figure}

In this case the noncommutative ring is simply $A=\End_R(R\oplus M)$. It can be decomposed in four pieces as follows:
\begin{equation}\label{eq:A-coni}
\begin{array}{cccccccc}
 A=&\Hom_R (R,R) &\oplus& \Hom_R(M,M) &\oplus& \Hom_R(R,M) &\oplus& \Hom_R(M,R)\\
 & \cong R && \cong R && \cong M && \cong M^\vee  \\
  & e_R && e_M && \alpha_i && \beta_i \\
\end{array}
\end{equation}
where in the last line we have written down the corresponding generators. The relevant morphisms are $\alpha_i$ and $\beta_i$, which satisfy the relations
\begin{equation}\label{eq:Fterms-coni}
\alpha_1\beta_i\alpha_2=\alpha_2\beta_i\alpha_1\ , \quad \beta_1\alpha_i\beta_2=\beta_2\alpha_i\beta_1\ , \quad i=1,2\ .
 \end{equation}
 These can be derived by checking the definition of $M$ in terms of the MF \cite{aspinwall-morrison}. However, the result can be repackaged as F-terms by defining a formal superpotential, which in this case happens to be (through no coincidence) the Klebanov--Witten one \eqref{eq:kw-superpotential}. $e_R$ and $e_M$ are the multiplicative identities (actually idempotents) of the ring at node $R$ and $M$ respectively. The quiver is depicted in Figure~\ref{fig:quiverConifoldNCCR}.
In this language, D-branes on the singular space are described as complexes of right $A$-modules, i.e. objects in the bounded derived category $D^b(\text{mod-}A)$. This makes sense because the bounded derived category of these modules is equivalent to the bounded derived category of coherent sheaves $D^b(\mathcal{X})$ on the resolved space \cite{Bondal:aa}. Moreover $A$-modules are equivalent to representations of the quiver. In particular, by studying the moduli space of the quiver representations corresponding to fractional D3-branes, we can recover the conifold variety $\mathcal{C}$. 

\begin{figure}[!ht]
\centering
\includegraphics[scale=1.25]{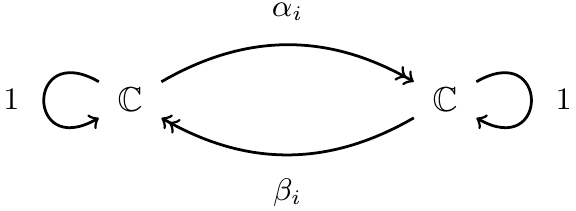}
\caption{$\vec{d}=(1,1)$ quiver representation of the conifold.}
\label{fig:quiverConifoldD3}
\end{figure}

A (finite-dimensional) quiver representation is defined by associating a (complex) vector space with each node of the quiver and a linear map with each arrow.\footnote{We will also encounter infinite-dimensional, but finitely-generated, quiver representations. These correspond to noncompact, or \emph{flavor}, branes.} $e_R$ and $e_M$ are set to the identity matrix $\mathbbm{1}$. The representation corresponding to a single D3-brane is shown in Figure~\ref{fig:quiverConifoldD3}, and is characterized by $\vec{d}=(1,1)$, collecting in a \emph{dimension vector} $\vec{d}$ the dimensions of the vector spaces at the two nodes.
The D3-brane splits into two fractional branes, one on each node. $\alpha_i$ and $\beta_i$ are complex scalar fields that transform in the bifundamental (and anti-bifundamental) of the product gauge group $\U(1)\times \U(1)$. Neglecting the decoupled diagonal $\U(1)$, we can say that these fields have charges $\pm 1$ under the relative $\U(1)$. 

The moduli space is parametrized by all the possible values of the maps $\alpha_i,\beta_i$  modulo the action of the relative $\U(1)$ gauge group. This naturally leads to the toric variety
\begin{equation}\label{eq:toric-coni}
\begin{array}{cccc}
 \alpha_1 & \alpha_2 & \beta _1 & \beta_2 \\ \hline 1 & 1 & -1 & - 1 \\ 
\end{array} \qquad\text{with}\qquad  |\alpha_1|^2+|\alpha_2|^2-|\beta_1|^2-|\beta_2|^2 = \xi\ ,
\end{equation}
where the last condition is the D-term of the $\U(1)$ gauge theory. We recognize this to be the toric description of the conifold space. When the parameter $\xi$ is zero, we have the singular conifold, while for $\xi\neq 0$ the space is resolved. The resolved phases correspond to $\xi>0$ and $\xi<0$.  We also see that in the first phase we will have the \emph{irrelevant ideal} condition $(\alpha_1,\alpha_2)\neq (0,0)$, while in the second phase we have $(\beta_1,\beta_2)\neq (0,0)$. By irrelevant ideal we mean the ideal associated with the excised locus, much like in the construction of $\pp^1$ from $\cc^2$ by excising its origin.

\subsubsection*{Stability}
\label{subsub:coni-stab}

Let us see here how to obtain the irrelevant ideal, when we consider the (resolved) space as the moduli space of a quiver representation. 
The quiver representation of interest here has dimension vector $\vec{d} = (1,1)$. In order to define the resolved ambient space, we first need to impose so-called \emph{stability conditions}. The first step requires assigning a vector $\vec{\theta}$ to this quiver such that $\vec{\theta} \cdot \vec{d}=0$, i.e. $\vec\theta=(-\xi,\xi)$. This gives us two choices up to rescaling: $\vec{\theta}_+:=(-1,1)$ and  $\vec{\theta}_-:=(1,-1)$. 

Before we can proceed, we should define the notion of \emph{subrepresentations}, and what it means for a representation to be destabilized by a subrepresentation. Physically this corresponds to our brane system decaying into a non-supersymmetric configuration. King's work \cite{king-git} then tells us how to insure that our $\vec{d}=(1,1)$ representation is stable in the appropriate way, and guarantees that stability is equivalent to satisfying D-term conditions in the gauge theory. 

One can show in detail how to exclude destabilizing subrepresentations for each of the phases $\vec\theta_\pm$. However, there is an easier and cleaner way of getting the answer \cite{wemyss-priv,Reineke:2015aa,engel-reineke}. Given our two-node quiver with a chosen $\vec{\theta}=(-\xi,\xi)$, one of the two nodes, say the left one, will have a negative $\vec\theta$-component, and the other one will have a positive component. Then, calling $V_\text{left}$ and $V_\text{right}$ the vector spaces associated to the two nodes, the \emph{semistable} representations are defined by the requirement that the space of paths from the ``negative'' to the ``positive'' node, i.e. from $V_\text{left}$ to $V_\text{right}$, must fully generate the space $\text{Hom}(V_\text{left}, V_\text{right})$ of the underlying vector spaces. For the phase $\vec{\theta}_+=(-1,1)$, this means that the $\alpha_i$ generate the right $\cc$, i.e. that they are not both zero. For the opposite phase, $\vec{\theta}_- = (1, -1)$, we must impose analogously that $\beta_1$ and $\beta_2$ are not both zero, so that they generate $\text{Hom}(V_\text{right}, V_\text{left})$. We have then found the two irrelevant ideal conditions for the two phases mentioned above. This trick will be used extensively in further sections.

\subsubsection*{The exceptional curve}
\label{subsub:coni-excP1}

The exceptional locus is a single, irreducible $\mathbb{P}^1$ curve. Let us explicitly see this. Given a choice of $\vec{\theta}=\vec{\theta}_\pm$, the moduli space of $\vec{d}=(1,1)$ representations corresponds to a resolution $\mathcal{X}_\pm$ of the singularity $\mathcal{C}$ we started with.
Therefore, there exists a blow-down map 
\begin{equation}\label{eq:exc-coni}
\pi_\pm: \ \mathcal{X}_\pm \longrightarrow \mathcal{C} : (\alpha_1, \alpha_2, \beta_1, \beta_2) \mapsto (x, y, z, t)\ ,
\end{equation}
where the coordinates of $\cc^4$ are functions of the quiver variables that are invariant under the toric $\cc^\ast$ corresponding to the complexification of the relative $\U(1) \subset \U(1) \times \U(1)$ of the quiver. Namely:
\begin{equation}\label{eq:gaugeinv-coni}
x+y=\alpha_1\beta_1\ , \quad x-y=\alpha_2\beta_2\ , \quad z =\alpha_1\beta_2\ ,  \quad t=\alpha_2\beta_1\ .
\end{equation}
These four paths generate all gauge-invariant functions of $\alpha_i,\beta_i$.

Here we are interested in the fiber over the origin, $\pi_\pm^{-1}(0)$. This means that we want the locus in $\mathcal{X}_\pm$ such that all gauge-invariants vanish. In the $\vec{\theta}_+$ phase this means that $\beta_1=\beta_2=0$, and $\alpha_1,\alpha_2$ parametrize a $\pp^1$.
In the $\vec{\theta}_-$ phase, the roles of $\alpha_i$ and $\beta_i$ are exchanged.

\subsubsection*{Following the length-one flop}
\label{subsub:coni-flop}

Let us briefly see how the relative $\U(1)$ gauge theory D-term in \eqref{eq:toric-coni} allows us to follow continuously the flop transition undergone by the exceptional $\pp^1$. 

For $\xi>0$ we see that $|\beta_1|^2=|\beta_2|^2=0$ and $|\alpha_1|^2+|\alpha_2|^2 = \xi$ gives a finite K\"ahler size two-sphere. As we let $\xi \to 0$, the size also goes to zero. Then, when $\xi<0$ we have $|\alpha_1|^2=|\alpha_2|^2 =0$, and $|\beta_1|^2+|\beta_2|^2=-\xi$ gives the size of a different sphere with opposite orientation.


\section{Laufer's example} 
\label{sec:lauf}

In this section we shall consider a class of pointlike threefold singularities that generalize the conifold studied before.\footnote{This has no relation to the so-called \emph{generalized conifold}, often seen in the topological string literature.} The generalization stems from the fact that such singularities are examples of length-two flops, as opposed to the length-one case.\footnote{Just as the length-one conifold threefold can be seen as a family of complex deformations of the $A_1$ twofold singularity over a complex plane \cite{atiyah}, the length-two Laufer's case can be seen as a family of deformed $D_4$ singularities over $\cc$ \cite{curto-morrison}.}

\subsection{The threefold flop}
\label{sub:lauf-3fold}

Consider the following equation in $\cc^7$ \cite{curto-morrison}:
\begin{equation}\label{eq:univ}
W_\text{univ}: x^2 + u y^2 +2 v y z + w z^2+(u w-v^2) t^2 = 0 \ .
\end{equation}
It describes a singular hypersurface with two small resolutions $W_\text{univ}^\pm$. This sixfold is called the \emph{universal flop of length two}. By definition, any threefold singularity $W_\text{threefold}$ that has a crepant resolution with a length-two curve as exceptional locus admits a morphism into \eqref{eq:univ}. More precisely, given the map $W_\text{threefold}\rightarrow W_\text{univ}$, the resolution $W_\text{threefold}^\pm \rightarrow W_\text{threefold}$ is the pullback of $W_\text{univ}^\pm\rightarrow W_\text{univ}$ (with $W_\text{univ}^+ \dashrightarrow W_\text{univ}^-$).


Old examples of length-two flops were provided by Laufer \cite{laufer} and Morrison--Pinkham \cite{pinkham}; these were later put into standard form by Reid \cite{reid}. 
In \cite{curto-morrison} it is described how to derive all length-two examples from the universal flop, and a $2 \ell \times 2 \ell = 4\times 4$ MF of the latter is given.\\

Let us now describe in more detail the class of singular flopping geometries studied by Laufer. Let $X$ be a rational, singular CY threefold. Let $\mathcal{X} \to X$ be a small resolution and $C \cong \pp^1$ the exceptional locus; let us call $\mathcal{N}$ the normal bundle to $C$ in $\mathcal{X}$. Then $\mathcal{N}$ must be a sum of line bundles $\mathcal{N} \cong \mathcal{L}_1 \oplus \mathcal{L}_2$, with $c(\det \mathcal{N}) = c_1(\mathcal{L}_1)+c_1(\mathcal{L}_2)$. By the adjunction formula, $T \mathcal{X} = TC \oplus \mathcal{N}$, one deduces that $c_1(\mathcal{L}_1)+c_1(\mathcal{L}_2) = -c_1(C)$, where $\int_C c_1(C) = 2$. Defining the Chern numbers $(n_1, n_2):= \big(\int_C c_1(\mathcal{L}_1 \big), \int_C c_1(\mathcal{L}_2) \big)$, we see that $n_1+n_2 = -2$. In order to have an isolated singularity, it turns out that we can only have $(n_1,n_2) = (-1,-1)$, $(-2,0)$, and $(-3,1)$, and this exhausts all possibilities \cite{laufer}. $(-1,-1)$ corresponds to the conifold flop, $(-2,0)$ to the so-called Reid's pagoda (i.e. $x^{2n}-y^2-tz=0 \subset \cc^4$ with $n \geq 2$), and $(-3,1)$ is the case of interest to this paper.

Indeed Laufer showed \cite{laufer} that, in the $(-3,1)$ case, $X$ can be written as a hypersurface inside $\cc^4$, and moreover we have a  family of such singularities labeled by an odd integer $k=2n+1 \geq 3$ (that is $n \in \mathbb{N}_{>0}$). The hypersurface has an isolated singularity at the origin of $\cc^4$, and its defining equation reads
\begin{equation}\label{eq:lauf}
W_\text{Laufer}: x^2 + y^3 + w z^2+w^{2n+1} y= 0\ \subset\ \cc^4\ .
\end{equation}
As predicted in \cite{curto-morrison}, \cite{aspinwall-morrison} explicitly showed that the above hypersurface is a specialization of \eqref{eq:univ}, obtained via the restriction
\begin{equation}
 t \to w^n\ , \quad u \to y\ , \quad v \to 0\ . \label{eq:univtolaufAM}
\end{equation}

\subsection{Matrix factorization and singular divisors}
\label{sub:lauf-MF}

Like in the conifold case, with each resolved phase is associated an MF of Laufer's threefold \eqref{eq:lauf}, i.e. the pairs $(\Phi_\text{L},\Psi_\text{L})$ and $(\Psi_\text{L},\Phi_\text{L})$ satisfying
\begin{equation}\label{eq:MFeqlauf}
\Phi_\text{L}\cdot  \Psi_\text{L} = \Psi_\text{L}\cdot  \Phi_\text{L} = W_\text{Laufer}\, \mathbbm{1}_{4\times 4}\ .
\end{equation}
Since the corresponding MF is known for the universal flop of length two, the restriction \eqref{eq:univtolaufAM} can be used to produce the two matrices for Laufer \cite{aspinwall-morrison}:
\begin{equation}\label{eq:MFlauf}
\Phi_\text{L} := \begin{bmatrix} x & -y&-z&-w^n \\ y^2 & x & w^n y & -z \\ w z & -w^{n+1} & x & y \\ w^{n+1} y & w z & -y^2 & x \end{bmatrix}\ , \quad 
\Psi_\text{L} := \begin{bmatrix} x & y & z & w^n \\ -y^2 & x & -w^n y & z \\ -w z & w^{n+1} & x & -y \\ -w^{n+1} y & -w z & y^2 & x \end{bmatrix} \ . 
\end{equation}
Notice that $\Phi_\text{L}=2 x\, \mathbbm{1}_{4\times 4} - \Psi_\text{L}$. The matrix $\Psi_\text{L}$ defines an MCM $R$-module $M= \coker \Psi_\text{L}$ through the exact sequence
\begin{equation}\label{eq:RRM-coni}
\begin{tikzcd}
0 \rar & R^{\oplus 4} \rar{\Psi_\text{L}} & R^{\oplus 4} \rar{} & M \rar &0\ .
\end{tikzcd}
\end{equation}
As done for the conifold case in \eqref{eq:det-coni}, we can extract families of Weil divisors directly from the MF \eqref{eq:MFeqlauf}. 
In the resolved space the rank-two module $M=\coker \Psi_\text{L}$ becomes a rank-two vector bundle. The divisors we are looking for are then Poincar\'e dual to the first Chern class of the bundle. The class can be determined as the locus where two of the bundle's (generic) sections become parallel. This locus can be identified already in the singular space, by requiring that two sections of $M$ be proportional to each other. 
To do this we use the isomorphism between $\coker \Psi_\text{L}$ and $\im \Phi_\text{L}$: When the domain of the map $\Phi_\text{L}$ is restricted to be $\coker \Psi_\text{L}$, the map is bijective (this is valid on generic points of the Laufer's threefold).\footnote{The space $\im\Phi_\text{L}$ is two-dimensional (when $\Phi_\text{L}$ is applied on $R^{\oplus 4}$); moreover when $\Phi_\text{L}$ acts on elements of the two-dimensional space $\im \Psi_\text{L}$ it gives zero, hence $\im \Psi_\text{L}\subset \ker \Phi_\text{L}$, which is also two-dimensional. Therefore $\im \Psi_\text{L}\cong \ker \Phi_\text{L}$. Hence, inside $\coker \Psi_\text{L}$, the kernel of $\Phi_\text{L}$ is empty and the map is invertible.}
Hence, the locus where two sections of $\coker \Psi_\text{L}$ are parallel is the same as the locus where two sections of $\im \Phi_\text{L}$ are parallel.
Since the image $\im \Phi_\text{L}$ is generated by the columns of $\Phi_\text{L}$, we can choose two columns of $\Phi_\text{L}$ and find the locus where these become parallel. 
Take e.g. the last two columns of $\Phi_\text{L}$. The locus we are looking for is given by
\begin{equation}
  \mbox{rank} \,\begin{bmatrix}
  -z& -w^n \\  w^n y & -z \\ x & y \\  -y^2 & x \end{bmatrix} \leq 1\ .
\end{equation}
When all the two-by-two minors vanish we obtain the vanishing locus of the following ideal:
\begin{equation}\label{LauferSectionGlob}
  \left(z^2 +w^{2 n} y\ ,\ -y z +w^n x\ , \   x z + w^n y^2\ ,\  \eqref{eq:lauf} \right)  \ .
\end{equation}
On the other hand, one can make a different choice, e.g. the second and last columns give
\begin{equation}\label{Locus42}
   \left(y z +  x w^{n}\ , \ -x y +w^{n+1} z\ , \ y^2 + w^{2n+1} \ , \  \eqref{eq:lauf} \right)\ ,
\end{equation}
while the first and last give
\begin{equation}\label{Locus41}
  \left(-x z+w^{n} y^2\ , \ x y + z w^{n+1}  \ , \  x^2 +w^{2n+1} y \ , \  \eqref{eq:lauf} \right) \ .
\end{equation}
In fact, taking generic combinations of columns and requiring them to be parallel gives a whole family of Weil divisors. We will explore this further at the end of Section~\ref{sec:nc-div}.\\

Interestingly, notice that Laufer's geometry \eqref{eq:lauf} can also be thought of as (a patch of) an elliptic fibration over a noncompact $\cc^2_{(w,z)}$ base. It is in fact described by the Weierstrass model (after a trivial redefinition $y \mapsto -y$)
\begin{equation}
x^2=y^3 + f(w,z)\,y+g(w,z)\ , \quad f:= w^{2n+1}\ , \quad g:= -wz^2 \ .
\end{equation}
The discriminant is $\Delta := 4f^3+27g^2= w^2 (4w^{6n+1}+27z^4)$. We notice that the seven-brane locus splits into one component with fiber type $II$ and one with fiber type $I_1$. At the intersection of the two loci, where the elliptic fibration is singular, the fiber type is $I_0^\ast$. This would na\"ively  correspond to a $D_4$ enhancement. However, we know that the resolved fiber over this locus only has one $\pp^1$. This can be blamed on the fact that the Kodaira classification is only reliable for elliptically-fibered K3 surfaces. 

However, this might be a hint that there is a T-brane effect at play, that is breaking this $D_4$ enhancement in a nonconventional way. An analogous situation was observed in an $\SU(5)$ F-theory setting in \cite{Esole:2011sm}: There, the Kodaira table na\"ively predicted an $E_6$-type fiber enhancement over a codimension-three locus, but the fiber type turned out to be entirely different (not even of $ADE$ type). The puzzle was resolved by \cite{Braun:2013cb}, who realized that the $E_6$ was partially broken due to a T-brane, or monodromic, effect.

This elliptic fibration has an extra (rational) section given, in the patch $w\neq 0$, by
\begin{equation}\label{EllSecLauf}
  x = \frac{z^3}{w^{3n}}\ , \quad\quad y = \frac{z^2}{w^{2n}} \ .
\end{equation}
In F-theory compactifications, extra sections of the elliptic fibration correspond to massless $\U(1)$ gauge symmetries in the lower-dimensional effective theory. The class of elliptic fibrations with one rational section has been studied by Morrison and Park in \cite{morrison-park}.  Laufer's threefold belongs to this class. Hence, F-theory on Laufer has a massless $\U(1)$ gauge boson. 
Locally, we can see that the extra section \eqref{EllSecLauf} is equivalent to the divisor~\eqref{LauferSectionGlob}.\footnote{Modulo a sign due to the redefinition of $y$.} It is then just one of the possible Weil divisors associated with the MF~\eqref{eq:MFlauf}. We thus see how the MF is related to the massless $\U(1)$'s in F-theory.
This observation, as well as the study of the local F-theory model provided by \eqref{eq:lauf} and its generalizations, will be the subject of a companion paper \cite{collinucci-fazzi-morrison-valandro}.


\section{Noncommutative crepant resolution} 
\label{sec:lauf-nccr}

\begin{figure}[!ht]
\centering
\includegraphics[scale=1.0]{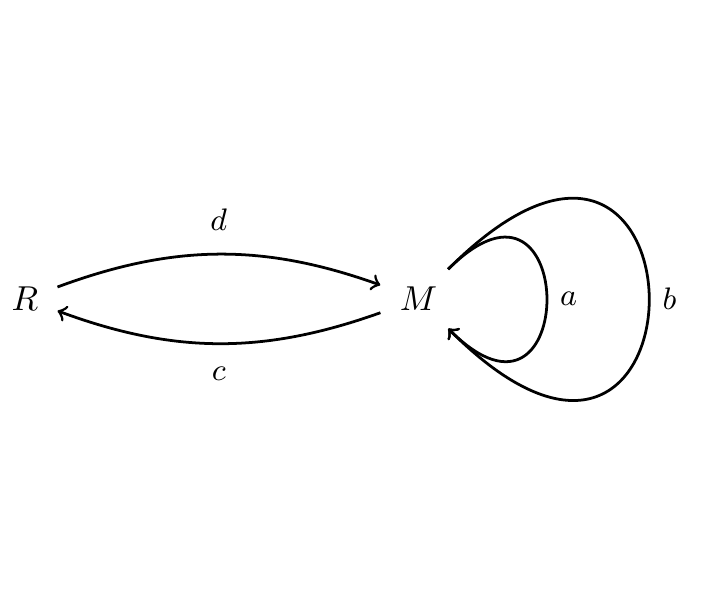}
\caption{Quiver for the NCCR of Laufer's example with $n=1$ \cite{aspinwall-morrison}. For $n>1$ one must add an extra loop at $R$.}
\label{fig:noncommquiver}
\end{figure}

In this section we will use the NCCR technique to extract the exceptional length-two $\pp^1$ locus of Laufer's singular threefold. Moreover, we will identify the class of divisors mentioned above in terms of quiver variables. We will use the generalization to three dimensions \cite{vdb-nccr} of classic results by Artin--Verdier \cite{artin-verdier} on $ADE$ twofold singularities to characterize the divisors as first Chern class of a vector bundle associated with an MCM module. Like in the conifold example, there are two inequivalent resolutions, associated with the two inequivalent MF's of Laufer's threefold and correspondingly with two MCM modules. For the NCCR we pick up one of the two MCM modules and we construct the noncommutative ring $A$ as follows:
\begin{equation}\label{eq:MCMMF1}
M= \coker \Psi_\text{L} \ , \quad A = \End_R(R \oplus M)\ .
\end{equation}
Without loss of generality, let us consider Laufer's example \eqref{eq:lauf} with $n=1$:
\begin{equation}\label{laufer1}
W_\text{Laufer}:\ x^2 +y^3 +w z^2 +w^3 y =0\ \subset\ \cc^4\ .
\end{equation}
This geometry will be our main case-study throughout the rest of the paper. The noncommutative resolution is described by the two-node quiver  depicted in Figure~\ref{fig:noncommquiver}.
The quiver has a path algebra, whereby linear combinations of arrows can be taken, and the product of two arrows is given by their concatenation, reading from right to left.\footnote{If the product does not correspond to a \emph{logical} concatenation, then it is zero. For instance, in this case, $d \circ a = 0$, whereas $a \circ d \neq 0$.} The path algebra must be supplemented with the following relations:
\begin{equation} \label{rels}
\mathcal{R}_\text{L} : \quad (b^2+dc) d = 0\ , \quad c (b^2+dc) =0 \ , \quad  ab+ba=0\ , \quad  a^2+b d c+ d c b+b^3=0\ .
\end{equation}
Like for the conifold, this data can be physically interpreted as the algebra of open strings attached to fractional branes that wrap the vanishing cycle of the geometry.

The maps (arrows of the quiver) $a,b,c,d$ can be expressed in terms of the MF data, as explained in \cite{aspinwall-morrison}.
In particular, any homomorphism $\alpha: R \rightarrow M$ lifts to a homomorphism  $\hat \alpha: R \rightarrow R^{\oplus 4}$ as follows:
\begin{equation}
\begin{tikzcd}
 & R \arrow[dl, "q"', dashed] \dar{\hat \alpha} && {R} \arrow[d, "\alpha"]\\
 R^{\oplus 4} \arrow[r, "\Psi_\text{L}"']  & R^{\oplus 4}  && {M}
\end{tikzcd}
\end{equation}
where $\hat \alpha \sim \hat \alpha + \Psi_\text{L} \cdot q$, for any $q: R \rightarrow R^{\oplus 4}$.

Conversely, any map $\beta: M \rightarrow R$ is expressible via its uplift to a map $\hat \beta: R^{\oplus 4} \rightarrow R$ as follows:
\begin{equation}
\begin{tikzcd}
 R^{\oplus 4} \arrow[r, "\Psi_\text{L}"] & R^{\oplus 4} \dar{\hat \beta}  && {M} \arrow[d, "\beta"]\\
 & R && R
\end{tikzcd}
\end{equation}
where we must impose the condition $\beta \cdot \Psi_\text{L} = 0$.

We can then follow \cite{aspinwall-morrison} and give the morphisms $a,b,c,d$ in terms of maps whose domain and codomain are either $R$ or $R^{\oplus 4}$:\footnote{We always consider cases where the quiver has two nodes, with the nontrivial node given by an MCM module $M=\coker \Psi$, with $\Psi$ an $n\times n$ matrix that factorizes the hypersurface equation, i.e. $(\Phi,\Psi)$ is a MF. The arrow from $R$ to $M$ are the morphisms from $R$ to $M$: These are generated by $n\times 1$ matrices with one entry equal to 1 and the other to 0. The morphisms from $M$ to $R$ are generated by the rows of $\Phi$ \cite{aspinwall-morrison}. There are also additional endomorphisms of $R$ and $M$ which sometimes need to be added. Typically, there are relations among these morphisms that allow to reduce the number of generators, leaving a smaller set of relations (this is what happens in Laufer's case).}

\begin{align}\label{eq:abcd-laufer}
a  &= \begin{bmatrix} 0 & 1 & 0 & 0 \\-y & 0 & 0 & 0\\ 0 &0 &0 &1\\ 0 & 0 & -y &0 \end{bmatrix} \ ,  &b =& \begin{bmatrix} 0 & 0 & 1 & 0 \\  0 & 0 & 0& -1 \\ -w &0 &0 &0\\ 0 & w & 0 &0 \end{bmatrix} \ ,\\
c &= \begin{pmatrix} x& -y & -z & -w \end{pmatrix} \ , &d=& \begin{pmatrix} 0 & 0 & 0 & 1 \end{pmatrix}^\text{t}\ .
\end{align}
From this, we can deduce relations between quiver loops (i.e. gauge-invariants) and affine $\cc^4$ coordinates. First, note that the following paths from left to right generate a vector space:
\begin{align} \label{lefttoright}
d &= \begin{pmatrix} 0 & 0 & 0 &1 \end{pmatrix}^\text{t}\,, \quad a d  = \begin{pmatrix} 0 & 0 & 1 & 0 \end{pmatrix}^\text{t}\,,\\ b d &= \begin{pmatrix} 0 & -1 & 0 & 0 \end{pmatrix}^\text{t} \,, \quad a b d = \begin{pmatrix} -1 & 0 & 0 & 0 \end{pmatrix}^\text{t} \ .
\end{align}
The affine ambient $\cc^4$ coordinates can be recovered in terms of loops on the left-hand side node:\footnote{Alternatively, $(x,y,z,w)$ can also be expressed as right-hand side loops \cite{aspinwall-morrison}. This is because $\Hom_R (R,R) \cong \Hom_R(M,M) \cong R$.}
\begin{equation} \label{coordsfromleft}
-c a b d = x\ , \quad c b d = y\ , \quad -c a d = z\ , \quad -cd = w\ .
\end{equation}
Other useful relations also hold, e.g.
\begin{equation}\label{eq:morerel-lauf}
c a^2 d = y w\ , \quad c b^2 d = w^2\ .
\end{equation}
Their usefulness will become clear in Section~\ref{sub:U1}.

\subsection{\texorpdfstring{Laufer as $4d$ $\mathcal{N}=1$ quiver gauge theory}{Laufer as 4d N=1 quiver gauge theory}}
\label{sub:lauf-4dN1quiv}

Having described the NCCR, which amounts to the path algebra of a quiver with relations, we will now extract the geometrically-resolved space from that data. In principle, we could already extract the hypersurface equation \eqref{laufer1} from the path algebra: The coordinate ring of the hypersurface is simply the center of the noncommutative ring (path algebra) $A= \text{End}_R (R \oplus M)$.

However, we will now use a more powerful technique that will not only allow us to reproduce the singular geometry, but will also give us its explicit resolutions in both its phases, and show us the flop transition. Mathematically, we will define a finite-dimensional representation of the quiver. This means that we replace the nodes with complex vector spaces, the arrows become linear maps between them, and the relations in \eqref{rels} must still be imposed. In our case, the appropriate representation will have dimension vector $\vec{d}=(1,2)$, that is $d_R=1, d_M=2$ on the left-hand side and right-hand side nodes, respectively. These dimensions correspond to the ranks of the MCM modules, respectively.

\begin{figure}[!ht]
\centering
\includegraphics[scale=1.25]{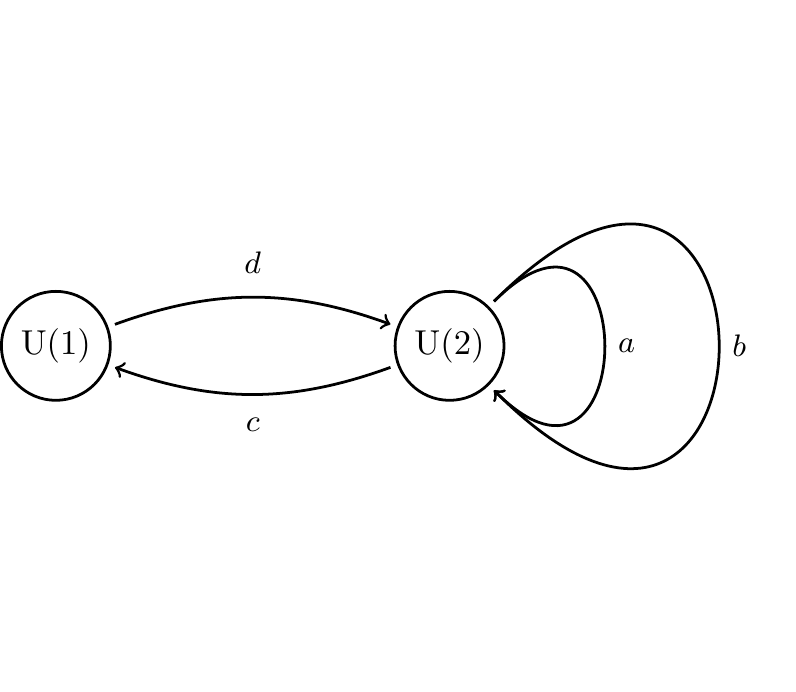}
\caption{Four-dimensional $\mathcal{N}=1$ quiver gauge theory for Laufer's example with $n=1$.}
\label{fig:quiver}
\end{figure}

Physically, we are studying the $\mathcal{N}=1$ quiver gauge theory that arises from probing the singularity with a spacetime-filling D3-brane. The theory has gauge group $\U(1) \times \U(2)$, and is depicted again in Figure~\ref{fig:quiver}.
The superpotential has been derived in \cite{aspinwall-morrison}, and reads:
\begin{equation}\label{eq:lauf-superpot}
\mathcal{W}_\text{L} = d c b^2+\tfrac{1}{2} c d c d + a^2 b+\tfrac{1}{4} b^4\ .
\end{equation}
The corresponding F-term relations are the following:
\begin{subequations}\label{eq:fterms-lauf}
\begin{align}
\partial_{c} \mathcal{W}_\text{L} &= (b^2+dc) d\,,\label{fterm1}\\  
\partial_{d} \mathcal{W}_\text{L} &= c (b^2+dc)\,, \label{fterm2}\\
\quad \partial_{a} \mathcal{W}_\text{L} &= ab+ba\,, \label{fterm3}\\
\partial_{b} \mathcal{W}_\text{L} &= a^2+b d c+ d c b+b^3\,. \label{fterm4}
\end{align}
\end{subequations}
For this theory, $a$ and $b$ are adjoint $\U(2)$ fields, i.e. two-by-two matrices. $d$ and $c$ are bifundamentals, s.t. $d$ is a two-by-one matrix, and $c$ a one-by-two matrix. The theory has a (classical) moduli space parametrized by the gauge-invariant combinations of these four fields. However, it can also be described by using coordinates of the quotient space
\begin{equation}
 \cc^{12} \left\langle a, b, c, d \right\rangle / \,(\cc^* \times \GL_2(\cc))\ ,
\end{equation}
where $\cc^*$ is the complexified $\U(1)$ gauge group, and $\GL_2(\cc)$ is the complexified $\U(2)$ gauge group. Note though, that just as projective spaces are quotients of an appropriately \emph{punctured} complex vector space, so must this $\cc^{12}$ be punctured. More precisely, we must exclude some algebraically closed subspaces before we can define the quotient. This per se does not insure that the result be nonsingular, but it at least enforces that the resulting space be Hausdorff. In gauge theory terms, this amounts to imposing D-term constraints of the following form:
\begin{equation} \label{generaldterm}
\sum_{a: \ast \rightarrow v}  \phi^a {\phi^a}^\dagger-\sum_{a: v \rightarrow \ast} {\phi^a}^\dagger \phi^a = \theta_v \, \mathbbm{1}_{N_v \times N_v}\ ,
\end{equation}
for each node $v$, whereby one must also impose 
\begin{equation}\label{eq:v-constr}
\sum_v \theta_v\, \dim V_v = 0\ , 
\end{equation}
$\dim V_v$ being the dimension of the vector space at the $v$-node (i.e. the entry $d_v$ of $\vec{d}$). Labeling our nodes $v=R,M$, this translates to the following conditions:
\begin{subequations}\label{eq:D-lauf}
\begin{align}
& c c^\dagger -d^\dagger d= \theta_R\ , \\
&d d^\dagger-c^\dagger c +[a, a^\dagger]+[b, b^\dagger] = \theta_M \, \mathbbm{1}_{2\times 2}\ .
\end{align}
\end{subequations}
Here, we must impose $\theta_R+2 \theta_M = 0$. 
Once we have properly excised the bad loci, and taken the quotient by the gauge group action, the moduli space we are left with (after imposing the relations \eqref{eq:fterms-lauf}) is expected to be the CY threefold probed by the D3-brane. A nonzero choice for the $\vec{\theta}=(\theta_R,\theta_M)$ vector corresponds to resolving the singular space.\\

Let us apply the same trick used to determine the stability of the $\vec{d}=(1,1)$ quiver representation of the conifold. In Laufer's case, the quiver representation of interest will have dimension vector $\vec{d} = (1,2)$. Again there are two choices of  a $\vec\theta$-vector satisfying $\vec{\theta} \cdot \vec{d}=0$ up to rescaling: $\vec{\theta}_+:=(-2,1)$ and  $\vec{\theta}_-:=(2,-1)$. We will study both separately, and then see how to make a smooth flop transition from one to the other. Like for the conifold, the semistable representations are defined by the requirement that the space of paths from $V_\text{left}$ to $V_\text{right}$ must fully generate the space $\text{Hom}(V_\text{left}, V_\text{right})$ of the underlying vector spaces. For the phase $\vec{\theta}_+=(-2,1)$, we must then impose that
\begin{equation} \label{eq:pathsLR}
\text{Paths}(R, M) = \langle d, a d, b d, a b d \rangle \cong \cc^2\ .
\end{equation}
In order for this to be true, we must require that $(d, a d, b d)$ not all be collinear, when viewed as column two-vectors. For the opposite stability condition, $\vec{\theta}_- = (2, -1)$, we must impose
\begin{equation}\label{eq:pathsRL}
\text{Paths}(M,R) = \langle c, c a, c b, c a b \rangle \cong \Hom(\cc^2, \cc) \cong \cc^2\ .
\end{equation}
This means that $(c, ca, cb)$ should not all be collinear as row two-vectors.

We can summarize the situation by constructing the two following irrelevant ideals  corresponding to the two resolution phases:
\begin{align} \label{irrelevantideals}
\vec{\theta}_+= (-2,1)\ :& \quad I_+ := (a d \wedge d, b d \wedge d) \ ; \\ 
\vec{\theta}_- = (2,-1)\ :& \quad  I_- := (c a \wedge c, c b \wedge c)\ .
\end{align}
In the phase $\vec{\theta}_+$ ($\vec{\theta}_-$), the elements of $I_+$ ($I_-$) cannot vanish simultaneously. We will later see that these ideals are made of the homogeneous coordinates of the exceptional $\pp^1$ in each phase.

\subsection{Finding the exceptional curve}
\label{sub:lauf-excP1}

In this section, we will find the fiber of the resolution in either phase, and see that it corresponds to a $\pp^1$. Before doing so, let us briefly explain some generalities.

Given a choice of $\vec{\theta}=\vec{\theta}_\pm$, the moduli space of $\vec{d}=(1,2)$ representations corresponds to a resolution $\mathcal{X}_\pm$ of the singular hypersurface $W_\text{Laufer}$ we started with (with $n=1$), i.e.
\begin{equation}\label{eq:lauf-n1}
W_\text{Laufer} : \ x^2 +y^3 +w z^2 +w^3 y =0 \ \subset\ \mathbb{C}^4\ .
\end{equation}
Therefore, there exists a blow-down map 
\begin{equation}\label{eq:blowdown}
\pi_\pm: \ \mathcal{X}_\pm \longrightarrow W_\text{Laufer} : (a, b, c, d) \mapsto (x, y, z, w)
\end{equation}
where the coordinates of $\cc^4$ are functions of the quiver variables that are invariant under the $\cc^* \times \GL_2(\cc)$ group. Put compactly:
\begin{equation}\label{eq:ambient-C4R}
(x, y, z, w) \in R [a, b, c, d]^{\cc^* \times \GL_2(\cc)}\ .
\end{equation}
These invariants were derived in \cite{aspinwall-morrison} and are reported in \eqref{coordsfromleft}. The important point is that they are made as traces of gauge-invariant loops. Now, we are interested in the fiber of the origin $\pi_\pm^{-1}(0)$. This means that we want the locus in $\mathcal{X}_\pm$ such that gauge-invariants are traceless. These are traces of either numbers or two-by-two matrices, and all their positive powers. Therefore, all loops must correspond to nilpotent maps, respecting the stability conditions \cite{engel-reineke, Reineke:2015aa}.

\subsubsection*{\texorpdfstring{Phase $\vec{\theta}_+ = (-2,1)$}{Phase theta+ = (-2,1)}}
\label{subsub:theta+}

Let us start with the phase $\vec{\theta}_+ = (-2, 1)$. The most basic loops based on the left-hand node are:
\begin{equation}\label{eq:L-node}
R: \ c d\ , \quad c a^n b^m d \quad \forall n, m\ , \quad \text{and products thereof}\ ,
\end{equation}
where $a^n b^m$ means any combination of (integer) powers of those arrows. Requiring their nilpotency means setting them to zero, since they are given by one-by-one matrices. On the right-hand node we have:
\begin{equation}\label{eq:R-node}
M: \quad d c\ ,\quad d a^n b^m c\ ,\quad  a^n b^m \ ,\quad \forall n, m\ ,  \quad \text{and products thereof}\ .
\end{equation}
Note that the first two sets of loops are automatically nilpotent if we set $cd=0$, as required by the nilpotency of the left-hand side loops.

Taking both sets of variables together, we see that the fiber is given by the following ideal:
\begin{equation} \label{eq:ideal-reineke-laufer}
\big( c d, a^2, b^2, c a d, c b d, c a b d \big)\ .
\end{equation}
Remember that, in this phase, $ad, bd$ and $d$ cannot be collinear. Hence the ideal $(cd, cad, cbd)$ immediately implies $c=0$. All in all, we can simplify the answer to
\begin{equation}\label{eq:P1+ideal-laufer}
I_{\pp^1_+} := \big(c, a^2, b^2, a b+b a \big)\ .
\end{equation}
Given this ideal, and the irrelevant ideal $I_+=(ad \wedge d, bd \wedge d)$, we can choose a convenient basis that allows us to ``see'' the $\pp^1$ more directly. Let us illustrate this. 

From the irrelevant ideal, we have that $a$ and $b$ cannot vanish simultaneously. 
First, we would like to prove that $a$ and $b$ are proportional. Say, without loss of generality, that $a \neq 0$ (the considerations we will make are symmetric in $a\leftrightarrow b$). Then it must have a nontrivial one-dimensional kernel, generated by a vector $v_a$. The relation $ab+ba$ tells us then that $ab\, v_a=0$. From this we deduce one of two possibilities:
\begin{description}
\item $b v_a \propto v_a$. This is impossible, since a nilpotent matrix cannot have nonzero eigenvalues.
\item $b v_a = 0$. This implies that $\ker a\subseteq \ker b$.
\end{description}
Similarly, we can prove that $\ker b \subseteq \ker a$, which implies that $\ker a\cong \ker b$. Now, the nilpotency of $a$ and $b$ gives us the following picture
\begin{align}\label{eq:imker-lauf}
\im a\subseteq\ &\ker a \nonumber \\
&\cong\\
\im b \subseteq \ &\ker b \nonumber
\end{align}
from which we infer that either $\im a \cong \im b$, or one of the two variables is zero. In either case, the conclusion is that $a \propto b$. Now we can use the $\SL_2(\cc) \subset \GL_2(\cc)$ symmetry to fix a basis for $\cc^2$ such that $ad$ and $bd \propto (1,0)^\text{t}$ and $d \propto (0,1)^\text{t}$. Using a combination of the left $\cc^\ast$ and the two $\cc^\ast$ inside $\GL_2(\cc)$, we can fix $d = (0,1)$. This leaves one $\cc^\ast$ that acts on the pair $(ad, bd)$ via rescaling. Summarizing, we have the following Ansatz for the most general solution to F and D-terms:
\begin{equation} \label{nilp1}
a = \begin{bmatrix} 0 & \alpha \\ 0 & 0 \end{bmatrix}\ ,\quad b = \begin{bmatrix} 0 & \beta \\ 0 & 0 \end{bmatrix}\ ,\quad d = \begin{pmatrix}0 \\1 \end{pmatrix}\ , \quad c=0\ ,
\end{equation}
with an action of the following $\cc^\ast$ subgroup of $\GL_2(\cc)$:
\begin{align}\label{eq:C*theta+}
\cc^\ast: (a, b) &\mapsto \begin{bmatrix} \lambda & 0 \\ 0 & 1 \end{bmatrix} (a, b) \begin{bmatrix} \lambda^{-1} & 0 \\ 0 & 1 \end{bmatrix}\ , \\
d &\mapsto d\ .
\end{align}
Clearly, this turns $(ad, bd)\wedge d$ into homogeneous coordinates:
\begin{equation}\label{eq:homcoordsP1theta+}
(a d, b d)\wedge d = (\alpha, \beta) \mapsto (\lambda \alpha, \lambda \beta)\ .
\end{equation}
The fact that this pair is exactly the irrelevant ideal makes this into an actual $\pp^1[\alpha:\beta]$.

\subsubsection*{\texorpdfstring{Phase $\vec{\theta}_- = (2,-1)$}{Phase theta- = (2,-1)}}
\label{subsub:theta-}

In this phase, $c, ca$ and $cb$ cannot all be collinear. Hence, the ideal $(cd, cad, cbd)$ implies $d=0$. The curve is given by
\begin{equation} \label{eq:P1-ideal-laufer}
I_{\pp^1_-} := \big(d, a^2, b^2, a b+b a \big)\ .
\end{equation}
On this side, the Ansatz for the $\pp^1$ is equally simple to find, and turns out to be the following:
\begin{equation} \label{nilp3}
a = \begin{bmatrix} 0 & \alpha \\ 0 & 0 \end{bmatrix}\ , \quad b = \begin{bmatrix} 0 & \beta \\ 0 & 0 \end{bmatrix}\ , \quad c = \begin{pmatrix}1 &0 \end{pmatrix}\ , \quad d=0\ .
\end{equation}
The gauge group is broken to the following $\cc^\ast$:
\begin{align}\label{eq:C*theta-}
\cc^\ast: (a, b) &\mapsto \begin{bmatrix} 1 & 0 \\ 0 & \lambda^{-1} \end{bmatrix} (a, b) \begin{bmatrix} 1 & 0 \\ 0 & \lambda \end{bmatrix}\ ,\\
c &\mapsto c\ .
\end{align}
The homogeneous coordinates of the $\pp^1$ are then
\begin{equation}
(c a, c b) \wedge c = (\alpha, \beta)\ ,
\end{equation}
which constitute the irrelevant ideal for this phase, and transform as expected.

\subsection{Following the flop transition}
\label{sub:lauf-flop}

In the previous two sections we showed how the resolved space looks by imposing two opposite stability conditions, $\vec{\theta} = (-2,1)$ and $(2, -1)$. In both cases, we see a $\mathbb{P}^1$. Hence, the flop transition has to do with negating the stability parameter. However, we would like to follow this transition continuously through the singularity, see a two-sphere shrink to zero size, and a new one grow. In order to do this it is best to use the D-term constraints instead of $\vec{\theta}$-stability. For reading convenience, we repeat these here. Taking $\vec\theta = (-2 \xi, \xi)$, we have:
\begin{align}
&d^\dagger d- c c^\dagger = 2 \xi\ , \\
&d d^\dagger-c^\dagger c +[a, a^\dagger]+[b, b^\dagger] = \xi \mathbbm{1}_{2\times 2}\ .
\end{align}
Let us make the following overarching Ansatz, which interpolates between \eqref{nilp1} and \eqref{nilp3}:
\begin{equation}\label{eq:nilp-inter}
a = \begin{bmatrix} 0 & \alpha \\ 0 & 0 \end{bmatrix}\ , \quad b = \begin{bmatrix} 0 & \beta \\ 0 & 0 \end{bmatrix}\ , \quad c = \begin{pmatrix} \sqrt{2} \gamma & 0 \end{pmatrix}\ , \quad d = \begin{pmatrix} 0 \\ \sqrt{2} \delta \end{pmatrix}\ , \quad \gamma \delta=0\ .
\end{equation}
It satisfies the F-term constraints \eqref{rels}. Plugging the above Ansatz into the D-term constraints, we get the following equations:
\begin{align}\label{eq:D-mix}
|\delta|^2-|\gamma|^2 &= \xi\ ,\\
\begin{bmatrix} |\alpha|^2+|\beta|^2-2 |\gamma|^2 &0 \\ 0 & 2 |\delta|^2-|\alpha|^2-|\beta|^2 \end{bmatrix} &= \xi \mathbbm{1}_{2\times 2}\ ,
\end{align}
with $\delta = 0$ when $\xi<0$ and $\gamma =0$ when $\xi>0$. We can rewrite this system as follows:
\begin{align}\label{eq:D-mix-fin}
&|\delta|^2-|\gamma|^2 = \xi\ , \\
&|\alpha|^2+|\beta|^2 = |\delta|^2+|\gamma|^2\ .
\end{align}
Now we see the transition continuously. Starting at $\xi >0$, our solution at the level of F-terms for the two-sphere dictates $\gamma=0$. Hence, we see that $|\delta|^2 = \xi$, which implies that
$|\alpha|^2+|\beta|^2 = \xi$, signaling a finite K\"ahler size for the exceptional sphere. As we send $\xi \rightarrow 0$, the size goes to zero. After transitioning to $\xi <0$, our solution for the sphere imposes $\delta=0$. Now we have $|\gamma|^2 = -\xi$, and hence $|\alpha|^2+|\beta|^2 = -\xi$, which corresponds to a different sphere of finite size.


\section{Weil divisors} 
\label{sec:nc-div}

In this section we will analyze in detail the Weil divisors associated with the small resolutions. Already for the conifold, we may see that they can be detected both in the singular phase and in the resolved phase. 

In M-theory geometric engineering, these divisors play an important role. The singularities we are studying can be obtained from a smooth hypersurface CY space by restricting its complex structure (specializing the defining equation). In this process, the threefold can gain {\it new} codimension-one submanifolds. In the singular space these are Weil divisors, that in the resolved phase become honest Cartier divisors. In M-theory, abelian gauge symmetries emanate from the reduction of the supergravity $C_3$-form along harmonic, normalizable two-forms that are Poincar\'e dual to the new divisors. In F-theory compactifications not all the divisors will correspond to abelian gauge symmetries. However,  these extra divisors are the natural objects to use in order to produce massless abelian gauge bosons in the effective theory. As we have seen in \eqref{EllSecLauf}, in this case the elliptic fibration develops an extra (rational) section that belongs to the family of Weil divisors; this is the condition for the F-theory to have an abelian gauge symmetry~\cite{Morrison:1996pp,morrison-park}.

In this section, we will introduce a new way of detecting such divisors in algebraic varieties that admit small resolutions. The conifold is an obvious case: It admits two families of Weil divisors $|D_\pm|$ whose union $|D_++D_-|$ is in the class of a Cartier divisor, but that are separately only Weil. In either resolution phase $\vec{\theta}_\pm$, the divisor $D_\pm$ will intersect the exceptional $\pp^1$ at a point, and $D_\mp$ will actually contain it as the total space of $\cO(-1)_{\pp^1}$ over it.

We will explore how to carry out this analysis for Laufer's example with $n=1$ in this section. We will find that, here too, there are two divisors $D_\pm$ such that one intersects the $\pp^1$ and the other contains it in one phase, and vice versa in the other phase. From the quiver perspective, we will actually recover the extra divisor, and the whole linear system in which it moves, as opposed to only the representative corresponding to the extra section.

\subsection{Divisors from the quiver}
\label{sub:divInNCCR}
\subsubsection*{\texorpdfstring{Divisor in phase $\vec{\theta}=(-2,1)$}{Divisor in phase theta=(-2,1)}}
\label{sub:divtheta+}

Let us start by finding the family of divisors $|D_+|$, which is the one that intersects the exceptional $\pp^1$ at a point in the phase $\vec{\theta}_+=(-2,1)$. The idea is to construct a line bundle such that the zero locus of its sections intersects the exceptional $\pp^1$ once. In M-theory, this will mean that we have a $\U(1)$ gauge group, and matter with charge one under it, given by a membrane wrapping the sphere. 

The construction is straightforward. First, note that our resolved space $\mathcal{X}_+$ comes equipped with a tautological bundle of the form 
\begin{equation}
\mathcal{V}_\text{taut} = \mathcal{L}_R \oplus \mathcal{V}_M\ ,
\end{equation}
where $\mathcal{L}_R$ is a line bundle whose structure group is the  left $\cc^\ast$, and $\mathcal{V}_M$ is a rank-two vector bundle with structure group identified with $\GL_2(\cc)$. The arrows in the quiver correspond to sections of these bundles as follows:
\begin{equation}
d \in \Gamma(\cL_R^\vee \otimes \cV_M)\ , \quad c \in \Gamma(\cL_R \otimes \cV_M^\vee)\ , \quad a, b \in \Gamma({\rm End}(\cV_M))\ .
\end{equation}
There is an ambiguity that allows us to twist $\mathcal{V}_\text{taut}$ into $\mathcal{V}_\text{taut} \otimes\tilde \cL$, where $\tilde\cL$ is any line bundle. This is equivalent to the statement that an overall $\U(1)$ gauge group decouples from the quiver theory. We can gauge fix this by tensoring with $\cL_R^{\vee}$, such that now
\begin{equation}\label{eq:Vtauttheta+}
\mathcal{V}_\text{taut} = \cO \oplus \cV\ ,
\end{equation}
where the first summand is the trivial bundle (i.e. the structure sheaf of $\mathcal{X}_+$), and the second a rank-two vector bundle.
Now, we have the following assignments to the various arrows:
\begin{equation}
d \in \Gamma(\cV)\ , \quad c \in \Gamma(\cV^{\vee})\ , \quad a, b \in \Gamma({\rm End}(\cV))\ .
\end{equation}
Artin--Verdier theory \cite{artin-verdier}, and its generalization to small-resolved threefolds by Van Den Bergh \cite{vdb-nccr,vdb-flops}, tells us how to construct a line bundle $\cL$ such that $c_1(\cL) = c_1(\mathcal{V}_\text{taut})$. In those references it is proven that the vector bundle $\mathcal{V}$ in \eqref{eq:Vtauttheta+} must occur in an exact sequence of the form
\begin{equation}\label{eq:exact1}
\begin{tikzcd}
0 \rar& \cO \rar & \cV \rar & \cL \rar &0\ ,
\end{tikzcd}
\end{equation}
where the divisor associated to $\cL$ intersect the exceptional $\pp^1$ once. To see that such a sequence must exist is simple. Pick a generic section of $\cV$, say $d$. This defines the first map $d: \cO \rightarrow \cV$. Since $d$ is nowhere vanishing, the rank of this map is always one. Therefore, the cokernel of the map must be a line bundle. Define $\cL$ to be that cokernel. The exterior product $\wedge\, d$ defines an explicit map $\wedge\, d: \cV \rightarrow \cL$ such that \eqref{eq:exact1} is an exact sequence:
\begin{equation}\label{eq:exact1-d}
\begin{tikzcd}
0 \rar& \cO \rar{\cdot\, d} & \cV \rar{\wedge \,d} & \cL \rar &0\ .
\end{tikzcd}
\end{equation}
This line bundle clearly satisfies $c_1(\cL) = c_1(\cV)$. From here, we see that we can generate all sections of $\cL$ as
\begin{equation} \label{genLplus}
\Gamma(\cL) = \langle ad \wedge d,\, bd \wedge d,\, a b d \wedge d \rangle\ .
\end{equation}
Note that, on the locus \eqref{eq:P1+ideal-laufer} of the $\pp^1$, the third generator $a b d \wedge d$ is identically vanishing. Comparing the remaining sections with the irrelevant ideal $I_+$ in \eqref{irrelevantideals}, we conclude that
\begin{equation}
\cL \cong \cO_{\pp^1}(1)\ .
\end{equation}
The irrelevant ideal we have excised imposes that $\cV$ be generated by its sections. This means that its restriction to the $\pp^1$ must decompose into a sum of line bundles of non-negative degree: $\cV|_{\pp^1} \cong \cO(a)\oplus \cO(b)$. The line bundle of equal first Chern class must then be $\Lambda^2 \cV|_{\pp^1} \cong \cO(a+b)$. Therefore,
\begin{equation}
\cV|_{\pp^1} \cong \cO(1)\oplus \cO\ .
\end{equation}
This is easily confirmed by looking at our Ansatz \eqref{nilp1}, which we repeat for convenience:
\begin{equation}
a = \begin{bmatrix} 0 & \alpha \\ 0 & 0 \end{bmatrix}\ , \quad b = \begin{bmatrix} 0 & \beta \\ 0 & 0 \end{bmatrix}\ , \quad d = \begin{pmatrix}0 \\1 \end{pmatrix}\ , \quad c=0\ .
\end{equation}
Since $ab=0$ on the $\pp^1$, $\cV|_{\pp^1}$ is generated by the sections $\langle d, ad, bd \rangle$. Given our Ansatz, these take the following form:
\begin{equation}\label{eq:ans-sec-theta+}
ad = \begin{pmatrix} \alpha \\0 \end{pmatrix}\ , \quad bd = \begin{pmatrix} \beta \\0 \end{pmatrix} \ , \quad d  = \begin{pmatrix} 0 \\1 \end{pmatrix}\ .
\end{equation}
These are clearly sections of $\cO(1) \oplus \cO$.

Finally, given the generators \eqref{genLplus} of the line bundle $\mathcal{L}$, our claim is that the $\U(1)$ corresponds to a family of divisors of the form 
\begin{equation} \label{d+quiver}
\sigma_{D_+} = c_1 (ad \wedge d)+c_2 (b d \wedge d)+c_3 (a b d \wedge d)\ .
\end{equation}

\subsubsection*{\texorpdfstring{Divisor in phase $\vec{\theta}=(2,-1)$}{Divisor in phase theta=(2,-1)}}
\label{sub:divtheta-}

Now we would like to describe the family of divisors $|D_-|$, intersecting the $\pp^1$ at a point in the $\theta_-$ phase. The logic is the same. We start by listing the sections of the dual bundle $\mathcal{V}^{\vee}$:
\begin{equation}\label{eq:dualsections}
\Gamma(\mathcal{V}^{\vee}) = \langle c, c a, c b, c a b \rangle\ .
\end{equation}
Pick a particular section, say $c$, and construct the exact sequence:
\begin{equation}
\begin{tikzcd}
0 \rar & \cO \rar{c \, \cdot} & \cV^{\vee} \rar{\wedge\, c} & \cL^{\vee} \rar & 0\ .
\end{tikzcd}
\end{equation}
In the phase $\vec{\theta}_-$, $c$ is nowhere vanishing, so the cokernel is a bona fide line bundle. From this, we can construct the generic $D_-$ divisor:
\begin{equation} \label{d-quiver}
\sigma_{D_-} = c_1 (c a \wedge c)+c_2 (c b \wedge c)+c_3 (c a b\wedge c) \ .
\end{equation}
To understand the fact that $|D_++ D_-|$ is a family of Cartier divisors, simply note that the product  $\sigma_{D_+} \sigma_{D_-}$ is a section of $\cL \otimes \cL^{\vee} \cong \cO$. Hence, such a section can be deformed by an arbitrary constant, thereby making the corresponding divisor miss the exceptional locus completely. This means that, under the blow-down map $\pi_\pm$ in \eqref{eq:blowdown}, this divisor can escape the singularity, and is therefore Cartier.

\subsection{Divisors as flavor branes}
\label{sub:flavored}

We now show how to obtain the divisor \eqref{d+quiver} by relying exclusively on the quiver gauge description of the singularity. We will add flavor nodes to the quiver, which will correspond to noncompact D7-branes along either Weil divisor. Naturally, the pointlike D3-brane (obtained from the $\vec{d}=(1,2)$ representation of the quiver), will have a flavor D3-D7 spectrum. We will then define the Weil divisors as the loci in the moduli space of the quiver where some of this matter becomes massless.
\begin{figure}[ht!]
\centering
\includegraphics[scale=1.25]{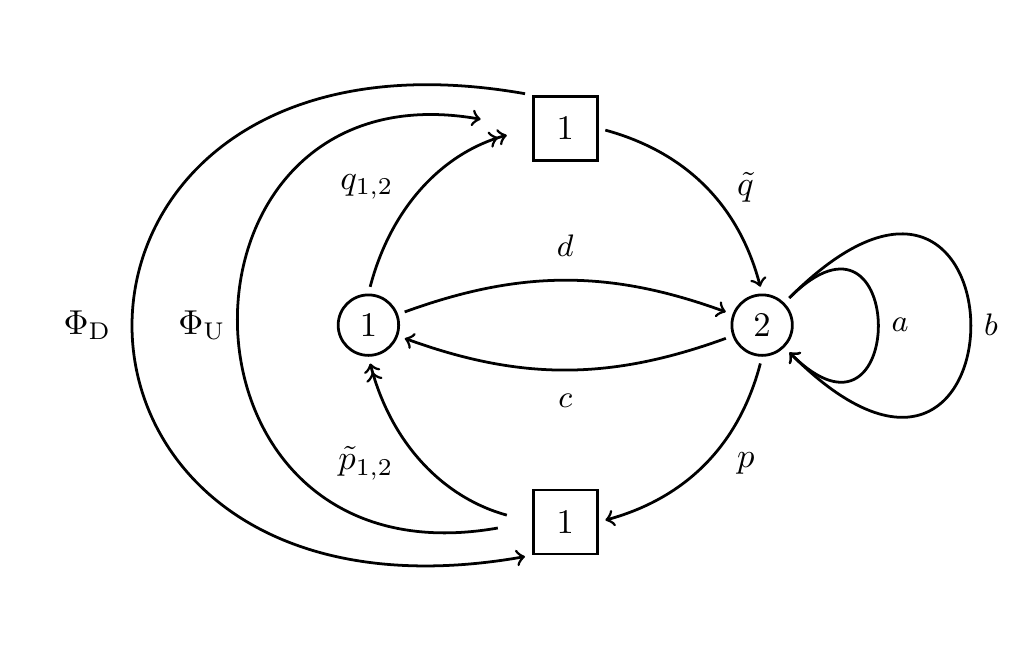}
\caption{Flavored version of Laufer's $n=1$ quiver.}
\label{fig:flavor-laufer}
\end{figure}%
Consider the flavored version of Laufer's quiver depicted in Figure~\ref{fig:flavor-laufer}. Square nodes correspond to flavor D7-branes, whereas round ones to fractional D3-branes. The arrows connecting color to flavor branes are D3-D7 states. Note that we must have two $q$ fields and two $\tilde{p}$ fields in order to make the flavor branes anomaly free.

The arrows connecting the boxes are D7-D7 states.
The superpotential \eqref{eq:lauf-superpot} will be modified by additional terms describing the interaction of these new states. Let us call $s_i$ the sections \eqref{eq:pathsLR} of the vector bundle $\cV$: $\Gamma(\cV)=\{s_i\}_{i=1}^4 := \{d,ad,bd,abd,\}$, and $\tilde{s}_i$ those of the dual bundle $\cV^\vee$ (i.e. \eqref{eq:dualsections}). The flavored superpotential can be schematically written as follows:
\begin{equation}\label{eq:lauf-flavorW}
\mathcal{W}_\text{L}^\text{flv} = \mathcal{W}_\text{L} + \mathcal{M}_n^\alpha\, p_\alpha\,  \tilde{p}_n  + \widetilde{\mathcal{M}}_{m\alpha}  \tilde{q}^\alpha\, q_m  + K \,\Phi_\text{U} p_\alpha \tilde{q}^\alpha + D^{mn}q_m \tilde{p}_n \Phi_\text{D}\ ,
\end{equation}
where 
\begin{equation}\label{mmtilde}
\mathcal{M}_n^\alpha :=  C^i_ns_i^\alpha\ , \qquad \widetilde{\mathcal{M}}_{m\alpha} := \tilde{C}_m^i\, \tilde{s}_{i\,\alpha}\ ;
\end{equation}
$\alpha=1,2$ is a $\U(2)$ index, $m,n=1,2$, $i=1,\ldots,4$, and $C_n^i,\tilde{C}_m^i,K,D^{mn}$ are numerical coefficients. 
The fields $\Phi_\text{U,D}$ describe the states between the two D7-branes. Let us consider the situation when their vev's vanish, i.e. no recombination between the two D7-branes takes place. 


We want to see what happens to the D3-D7 states when we move the D3-brane away from the singularity, that means when we give a generic vev to the sections $s_i$ and $\tilde{s}_i$ (satisfying the F-terms (\ref{fterm1}-\ref{fterm4})).
The masses of these states are described by the two-by-two matrices $\mathcal{M}$ and $\widetilde{\mathcal{M}}$ in \eqref{mmtilde}. 
At a generic point away from the singularity, both $\mathcal{M}$ and $\widetilde{\mathcal{M}}$ have maximal rank and all the D3-D7 states become massive. This means that the D3-brane is not on top of the D7-brane. On the other hand, when one of the two matrices has rank one (i.e. either $\det \mathcal{M} =0$ or $\det \widetilde{\mathcal{M}} =0$), a vector-like pair becomes massless: We interpret these states as the D3-D7 states that become massless when the D3-brane is on top of one of the two D7-branes. 

Notice that the columns of $\mathcal{M}$ ($\widetilde{\mathcal{M}}$) are two generic sections of the bundle $\cV$ ($\cV^\vee$). Hence $\det \mathcal{M} =0$ ($\det \widetilde{\mathcal{M}} =0$)
happens exactly when two sections of the bundle $\cV$ ($\cV^\vee$) become parallel, 
i.e. on top of the Weil divisors studied in Section~\ref{sub:divInNCCR}. 

We have then proven that the Weil divisors correspond to the locus where a D3-D7 state becomes massless.
To see this more concretely, take the locus 
$\det \mathcal{M} =0$. 
By field redefinition one can set $C_1^is_i=d$ (and the second column of $\mathcal{M}$ is a linear combination $ad,bd,abd$). The determinant of $\mathcal{M}$ then gives exactly the locus \eqref{d+quiver} obtained before. Analogous considerations hold for the locus $\det\widetilde{\mathcal{M}} =0$.

\subsection{Singular description}
\label{sub:U1}

Having described the two families of divisors $|D_+|$ and $|D_-|$ in the resolved geometries, we would like to describe the image of this family under the blow-down map \eqref{eq:blowdown}, in order to gain direct access to it in the singular space. In M-theory compactified on the CY threefold, the $\U(1)$ gauge group associated with these divisors exists, whether we resolve the singularity or not. Hence, the divisors we found must exist on the singular space as Weil divisors. Our strategy is to construct singlets out of sections $\sigma_{D_+}$ and $\sigma_{D_-}$ described in Section~\ref{sub:divInNCCR}, so that we can gain a description in terms of the affine $\cc^4$ coordinates $(x, y, z, w)$. 

Our strategy will be slightly counterintuitive. In order to describe the family $|D_+|$, which are the divisors that intersect the $\pp^1$ at a point in the $\vec\theta_+$ phase, we will actually first describe it in the $\vec\theta_-$ phase, and vice versa for the other divisor family. Here is why. In the $\vec{\theta}_-$ phase, we have an irrelevant ideal $I_- = (c a \wedge c, c b \wedge c)$. Incidentally, the generators of this ideal are sections of $\cL^{\vee}$, which is exactly what we need to create singlets from $\sigma_{D_+} \in \Gamma(\cL)$. The fact that they form an irrelevant ideal means that, if we create the ideal of products $(\sigma_{D_+} I_-)$, we will get an ideal whose zero locus must correspond to the zero locus $Z(\sigma_{D_+})$ of $\sigma_{D_+}$. For technical reasons, we will find it more convenient to throw in the singlets made by multiplying also by $(c a b \wedge c)$. The ideal in the resolved space will have the same zero locus; however, in the singular space, we will obtain a cleaner description of the Weil divisor family. Explicitly, define the following singlet matrix:
\begin{equation}\label{eq:Z+}
Z(\sigma_{D_+}) = Z\big(\sigma_{D_+} c a \wedge c,\ \sigma_{D_+} c b \wedge c ,\  \sigma_{D_+} c a b \wedge c\big)\ .
\end{equation}
The ideal $I_{+}$ can now be written entirely in terms of affine $\cc^4$ coordinates by using the relations in \eqref{coordsfromleft}. 
First, let us use the general fact that, given sections $s_i$ of $\cV$ and  $\tilde{s}_i$ of $\cV^{\vee}$, we have the following identity:
\begin{equation}
(s  \wedge r) (\tilde s \wedge \tilde r) = (s \cdot \tilde s) (r \cdot \tilde r) - (s \cdot \tilde r) (r\cdot \tilde s) \ .
\end{equation}
For instance, we would have identities like
\begin{equation}
 (c a \wedge c) (a d \wedge d) = (c a^2 d) (c d) - (c a d)^2 = - y w^2-z^2\ .
\end{equation}
Let us setup a matrix of such relations:
\begin{align}
Z(\sigma_{D_+}) &=\left[\begin{pmatrix}  c a \\ c b \\ c a b \end{pmatrix}  \wedge c\right] \otimes \left[ \begin{pmatrix} a d& b d & a b d \end{pmatrix} \wedge d \right] \nonumber \\
&= (cd) \begin{bmatrix} ca^2d & c a b d & ca^2 b d\\
c b a d & c b^2 d  & cb a b d \\ -c a^2 b d & c a b^2 d & -c a^2 b^2 d  \end{bmatrix} - \begin{pmatrix} cad \\ c b d \\ c a b d \end{pmatrix} \otimes \begin{pmatrix} cad & cbd & cabd \end{pmatrix} \ .
\end{align}
Plugging in \eqref{coordsfromleft} and \eqref{eq:morerel-lauf}, one obtains
\begin{align}
Z(\sigma_{D_+}) 
& =-w \begin{bmatrix} y w & -x & -y^2 \\ x & w^2 & -z w \\ y^2 & z w & y w^2 \end{bmatrix} - \begin{pmatrix} -z \\ y \\ -x \end{pmatrix} \otimes  \begin{pmatrix} -z & y & -x \end{pmatrix}\ .
\end{align}
Therefore, a generic section $\sigma_{D_+}$ will have the following description as a Weil divisor:
\begin{equation} \label{eq:Z+k}
Z(\sigma_{D_+}) \cdot \vec{k} = \begin{bmatrix} - yw^2-z^2 & x w+y z & w y^2-x z \\ -x w+zy & -w^3-y^2 & z w^2+x y\\ -y^2 w-x z & -z w^2+ x y & -y w^3 - x^2 \end{bmatrix} \cdot \begin{pmatrix} k_1\\k_2\\k_3 \end{pmatrix} = 0\ .
\end{equation}
We recover the locus \eqref{LauferSectionGlob} (with $n=1$) by taking $\vec{k} = (1,0,0)^\text{t}$. Notice that this coincides with the extra section \eqref{EllSecLauf} of the elliptic fibration given by the local F-theory model $W_\text{Laufer} = 0 \subset \pp^{231}[-y:x:1] \times \cc^2_{(w,z)}$. By setting $\vec{k} = (0,1,0)^\text{t}$ or $\vec{k} = (0,0,1)^\text{t}$ we obtain respectively the loci \eqref{Locus42} or \eqref{Locus41}.

Let us write down the singular description of the Weil divisor corresponding to $\sigma_{D_-}$ by going to the $\theta_+$ phase, multiplying by a matrix made from the irrelevant ideal in that phase. There is nothing to calculate, we simply transpose the three-by-three matrix we just constructed. Therefore, a generic section $\sigma_{D_-}$ will have the following description as a Weil divisor:
\begin{equation}\label{eq:Z-k}
Z(\sigma_{D_-}) \cdot \vec{k} = \begin{bmatrix} -y w^2-z^2 & -x w+y z & -w y^2-x z \\ x w+z y & -w^3-y^2 & -z w^2+x y\\ y^2 w-x z & z w^2+ x y & -y w^3 - x^2 \end{bmatrix} \cdot \begin{pmatrix} k_1\\k_2\\k_3 \end{pmatrix} = 0\ .
\end{equation}
This is the same family one would obtain from the MF $(\Psi_\text{L},\Phi_\text{L})$ and the corresponding MCM module $M^\vee=\coker \Phi_\text{L}$, following the computations of Section~\ref{sub:lauf-MF}.

\section{Higher-charge states}
\label{sec:highercharge}

Throughout the paper we have used the branes at singularities paradigm, whereby we describe the quiver gauge theory arising from a spacetime-filling D3-brane that is pointlike in the internal space. We will now switch to the so-called \emph{geometric engineering} picture in IIA, whereby we study the effective field theory that arises by reducing type IIA supergravity on our threefold, supplemented by the D-particles arising from various D2-branes wrapping the exceptional $\mathbb{P}^1$.\\

So far, Laufer's geometry seems to behave in perfect analogy with the conifold: It has a vanishing $\mathbb{P}^1$ that can be blown-up crepantly; the curve can be flopped; the threefold admits two Weil divisors, one of which cuts the curve at one point in one resolution phase, and the other one cuts the curve in the other phase. It seems then that the only novelty of this geometry is simply that it is more complicated to describe.

In this section, we will discover an important qualitative difference: IIA on Laufer's example admits hypers of higher charge under the $\U(1)$ generated by the Weil divisors. This is in stark contrast to the conifold, which only admits a hyper of charge one.

Once again we will use the example of the conifold as a familiar reference, in order to set the stage. In that case the two simple representations $\vec{d}=(1,0)$ and $\vec{d}=(0,1)$ correspond to the objects $\mathcal{O}_C$ (a D2-brane) and $\mathcal{O}_C(-1)[1]$ (an anti-D2 with flux) respectively, in the category $D^b(\mathcal{X})$. In the case of Laufer's example, it was explained in \cite{aspinwall-morrison} that $\vec{d}=(0,1)$ corresponds to $\mathcal{O}_C(-1)[1]$, whereas $\vec{d}=(1,0)$ corresponds to some different (not locally free) object $\mathcal{S}$ with support over the curve $C$. The following simple equation
\begin{equation}
\vec d= (1,2) = (1,0)+ 2\, (0,1)
\end{equation}
shows us that, at the K-theory level, the class of a point $p \in \mathcal{X}$ (i.e. a pointlike D0-brane) is given by  
\begin{equation}
\left[ \mathcal{O}_p \right] = \left[ \mathcal{S} \right] + 2\left[ \mathcal{O}_C(-1)[1]\right] = \left[ \mathcal{S} \right] - 2\left[ \mathcal{O}_C(-1)\right]\ \Leftrightarrow\ \left[ \mathcal{S} \right] = \left[ \mathcal{O}_p \right] + 2\left[ \mathcal{O}_C(-1)\right]\ .
\end{equation}
From this, we can conclude that the $\vec{d}=(1,0)$ representation must correspond to some bound state of two D2-branes. Hence, if we construct a hyper in the effective theory by taking a $\vec{d}=(0,1)$ D-particle (i.e. the anti-D2 with flux wrapped on $\pp^1$) and its corresponding anti-particle (i.e. the oppositely-oriented membrane corresponding to the object shifted by $[1]$), and normalize its charge to one, then we can also build a hyper of charge two from the $\vec{d}=(1,0)$ representation and its corresponding anti-brane.
To summarize in a succinct language:
\begin{align}
&\big( (0,1)\,, (0,1)[1] \big) \quad \longleftrightarrow \quad \text{hyper of charge one}\ , \\
&\big( (1,0)\,, (1,0)[1] \big) \quad \longleftrightarrow \quad \text{hyper of charge two}\ .
\end{align}
Constructing the anti-branes of a given representation can be done either by passing to the derived category of quiver representations $D^b(\text{mod-$A$})$ or, in the resolved geometry, to the derived category of coherent sheaves $D^b(\mathcal{X})$.

Now, these two hypers are never simultaneously massless. Depending on the value of the B-field, only one of them can be made massless at a time. The mass formula for a D-particle is given by the modulus $|Z|$ of its \emph{central} charge  
in four-dimensional $\mathcal{N}=1$ language. In turn, the central charge is a function $Z(\Gamma, B+i J)$ of its Ramond--Ramond charge vector $\Gamma$, and of the complexified K\"ahler modulus of the threefold. It gets worldsheet instanton corrections that are subleading at large volume, where the formula reduces to the following:
\begin{equation}
Z(\Gamma, B+i J) = -\int_{X_3} \Gamma \wedge e^{-(B+iJ)}+ \text{worldsheet instanton corrections} \ .
\end{equation}
In the case where the CY has no compact four-cycles (our case), the formula receives no $\alpha'$ corrections (as we will show momentarily), and simplifies to the following integral:
\begin{equation}
Z = \int_{\mathbb{P}^1} B+iJ-F\ ,
\end{equation}
where $F$ is the worldvolume DBI flux on the membrane. Now, our $(1,0)$ and $(0,1)$ hypers are branes that differ in rank, and in their DBI flux, by one unit of induced D0 charge. Hence, by shifting the B-field accordingly, either one can be made massless.

As promised, let us briefly argue that this formula for the central charge is indeed uncorrected. (This is a recurring folk theorem that we have been aware of for a long time.) The line of reasoning goes as follows. In general for, say, a one-modulus CY there are four periods of the $\Omega_3$-form of the mirror CY which, in some basis, would take the following form (see \cite{Aspinwall:2004jr} for an introduction):
\begin{align}
\Phi_0 &= P_0(z) \ ,\\
\Phi_1 &= \tfrac{1}{2 \pi i}\Phi_0 \log(z) + P_1(z)\ ,\\
\Phi_2 &= P_2(z)  \log^2(z) + P_2'(z)\ ,\\
\Phi_3 &= P_3(z)  \log^3(z) + P_3'(z)\ ,
\end{align}
where $z$ is the complex structure modulus of the mirror CY, and the $P_i(z)$ are polynomials in $z$. Now, $\Phi_0$ can be argued to be the central charge of the mirror to the D0-brane. The quotient $Q = \Phi_1/\Phi_0$ has a monodromy $Q \rightarrow Q+1$ around $z=0$. At large volume, with complex K\"ahler modulus $t = B+i J$, one makes the following match with the mirror side as follows:
\begin{align}
&\int_\text{two-cycle} t = \Phi_1/\Phi_0\nonumber \ , \\
&\int_\text{four-cycle} t^2 = \Phi_2/\Phi_0\ , \\
&\int_\text{six-cycle} t^3 = \Phi_3/\Phi_0\nonumber \ .
\end{align}
The monodromy above is mapped to the monodromy around the large volume point as $B \rightarrow B+1$.

If we had nontrivial four and six-cycles, then it would be nontrivial to solve for $t$ in terms of $z$. However given that in our case there is only a two-cycle, the only relation to satisfy is the following:
\begin{equation}
\int_{\pp^1} t = \frac{1}{2 \pi i} \log(z) + \frac{P_1(z)}{P_0(z)}\ ,
\end{equation}
which allows one to define a mirror map $z(t)$, which would typically be a nonperturbative series in $\alpha'$.
We can reabsorb the disturbing $P_1(z)$ piece by redefining $z \rightarrow \tilde z$ such that 
\begin{equation}
\int_{\pp^1} t = \frac{1}{2 \pi i} \log(\tilde z)\ .
\end{equation}
From this, we conclude that the exact central charge for a (B-type) D-brane is given by
\begin{equation}
Z(\Gamma, B+i J) = -\int_{X_3} \Gamma \wedge e^{-t}\ ,
\end{equation}
which reduces to 
\begin{equation}
Z = \int_{\pp^1} t-F
\end{equation}
in the case of a membrane. Clearly, given a value for $F$, there will be a point in moduli space where $Z=0$, giving rise to a massless state.

\section{Conclusions} 
\label{sec:conc}

In this paper we have studied in full detail a prototype CY threefold singularity that admits a flop of length two. Such geometries were of course previously known in mathematics, but only in terms of their algebraic properties. 

The approach we used here is to examine the worldvolume theory of a D3-brane probing Laufer's singularity in type IIB string theory. We have studied the resolution of the singular geometry by using its noncommutative crepant resolution, and subsequently its presentation as a quiver representation, i.e. by using the quiver geometric invariant theory method. 

We have shown how to extract the exceptional $\pp^1$ locus from the quiver representation, and how to follow continuously the flop transition the $\pp^1$ undergoes by looking at the four-dimensional $\mathcal{N}=1$ gauge theory D-terms. Moreover we have defined and studied (both in the singular and resolved phase) two families of Weil divisors of the geometry, which can be given a natural interpretation as $\U(1)$ divisors if we define a local F-theory model on Laufer's example.

By relying on the geometric engineering perspective, we also showed that IIA compactified on this class of geometries consists in a four-dimensional $\mathcal{N}=2$ $\U(1)$ gauge theory with at least one hypermultiplet of charge one, and one hypermultiplet of charge two. Both hypers can become massless, but at different points in the complexified K\"ahler moduli space.

We would now like to speculate on the possibility of having states of charge higher than two. These could be created if several fractional branes form appropriate bound states. In geometries like the conifold, these are usually ruled out. The main reason is the fact that the normal bundle $\mathcal{N}$ to the exceptional $\pp^1$ curve is strictly negative, and, since any bound state requires giving a vev to its sections, the states are not supported. However, flops of length two have $\mathcal{N} = \mathcal{O}_{\pp^1}(-3)\oplus \mathcal{O}_{\pp^1}(1)$, which means it is definitely conceivable to form bound states. We leave this interesting question for future work.

Finally, let us briefly comment on the F-theory interpretation of Laufer's example. We have seen that the latter geometry is already in Weierstrass form. The family of noncompact divisors we have found by our methods includes one divisor that can be regarded as an extra section to that fibration, implying the rank of the Mordell--Weil group is one. The exceptional $\pp^1$ represents a fiber enhancement over a codimension-two locus in the base of the fibration, giving rise to a charged hyper in six dimensions.


\section*{Acknowledgments} 
We would like to thank R.~Argurio, M.~Del~Zotto, I~Garc\'ia-Etxebarria, J.~J.~Heckman, H.~Jockers, J.~Karmazyn, D.~R.~Morrison, M.~Reineke, and M.~Wemyss for enlightening discussions, and especially D.R.M.~for collaboration on a related project. A.C.~and M.F.~thank the Banff International Research Station ``Geometry and Physics of F-theory'' workshop for a stimulating environment and hospitality during the final stages of this work. A.C.~is a Research Associate of the Fonds de la Recherche Scientifique F.N.R.S. (Belgium). The work of A.C~is partially supported by IISN - Belgium (convention 4.4503.15). The work of M.F.~is partially supported by the Israel Science Foundation under grant No.~1696/15 and 504/13, and by the I-CORE Program of the Planning and Budgeting Committee. The work of R.V.~has been supported by the Programme ``Rita Levi Montalcini for young researchers'' of the Italian Ministry of Research. 



\appendix 

\section{Morrison--Pinkham's example} 
\label{app:mpi}

Another example of threefold flop of length two was constructed by Morrison--Pinkham~\cite{pinkham}. The hypersurface singularity in $\cc^4$ is given by:
\begin{equation}\label{eq:mpi}
W_\text{MP} : \ x^2 +y^3 +w z^2 +w^{3} y  - \lambda w y^2-\lambda w^4=0 \ , \quad \lambda \in \cc\ .
\end{equation}
This is not a one-parameter family of singularities. There are only two distinct classes thereof, for $\lambda=0$ and $\lambda \neq 0$ \cite{aspinwall-morrison}. In particular it is easy to notice that the $\lambda = 0$ class is equivalent to the $n=1$ case \eqref{laufer1} of Laufer's example. The above threefold is a special case of the universal threefold flop \eqref{eq:univ}, obtained via the restriction \cite{aspinwall-morrison}
\begin{equation}\label{eq:univtompi}
t \to  -w \ , \quad u \to y -\lambda w \ , \quad v \to 0\ ,
\end{equation}
as one can easily check. We also obtain an MF of \eqref{eq:mpi} by applying \eqref{eq:univtompi} to the MF of $W_\text{univ}$ provided in \cite{aspinwall-morrison}. Explicitly:
\begin{subequations}\label{eq:MFmpi}
\begin{align}
&\Phi_\text{MP} = \begin{bmatrix} x & -y&-z&w\\ (y-\lambda w)y & x & -(y-\lambda w)w & -z \\ wz & w &x&y \\ -(y-\lambda w)w^{2} & wz & -(y-\lambda w)y & x\end{bmatrix}\ , \\
&\Psi_\text{MP} = \begin{bmatrix} x & y&z&-w\\ -(y-\lambda w)y & x & (y-\lambda w)w & z \\ -wz & -w &x&-y \\ (y-\lambda w)w^{2} & -wz & (y-\lambda w)y & x \end{bmatrix}\ ,
\end{align}
\end{subequations}
satisfying 
\begin{equation}\label{eq:MFeqmpi}
\Phi_\text{MP}\cdot  \Psi_\text{MP} = \Psi_\text{MP}\cdot  \Phi_\text{MP} = W_\text{MP}\, \mathbbm{1}_{4\times 4}\ .
\end{equation}
The quiver with relations producing an NCCR of \eqref{eq:mpi} was obtained in \cite{aspinwall-morrison}, and is equivalent to Laufer's one (depicted in Figure~\ref{fig:noncommquiver}). The relations in the path algebra read instead
\begin{equation} \label{rels-mpi}
\mathcal{R}_\text{MP}:\ (b^2+dc) d=0\ , \ c (b^2+dc)=0\ , \  ab+ba=0\ , \  a^2+b d c+ d c b+\lambda b^2 + b^3=0\ ,
\end{equation}
and can be obtained as the F-terms (i.e. cyclic derivatives) of the following superpotential \cite{aspinwall-morrison}:
\begin{equation} \label{eq:superpot-mpi}
\mathcal{W}_\text{MP} = d c b^2+\tfrac{1}{2} c d c d + a^2 b+\tfrac{1}{3} \lambda b^3 + \tfrac{1}{4} b^4\ .
\end{equation}
We will now show that the Ansatz \eqref{nilp1} also holds in the Morrison--Pinkham case, and can be obtained directly from the ideal
\begin{equation}\label{idealGenP1}
\big( c d, a^2, b^2, c a d, c b d, c a b d, (ab)^2 \big)\ ,
\end{equation}
where we just added the last relation with respect to the ideal \eqref{eq:ideal-reineke-laufer} (which is appropriate for Laufer's example). In that case, this relation was implied by the F-terms once one assumed the others. With the F-terms \eqref{rels-mpi}, we need to impose it by hand.

We will now show that the ideal \eqref{idealGenP1} implies that the nilpotent matrices $a$ and $b$ are proportional to each other. This is surely true if $a$ or $b$ are equal to zero. So we need to prove it for $a\neq 0$ and $b\neq 0$. We first show that if a nilpotent two-by-two matrix is nonzero, then its kernel and its image are isomorphic. In fact, if the matrix $m$ is nonzero ($m\neq 0$) and nilpotent ($m^2=0$), the dimension of  $\ker m$ and $\im m$ are equal to one. Moreover, since it is nilpotent, $\im m\subseteq \ker m$, implying $\im m\cong \ker m$ as the two subspaces are one-dimensional.

Now, we take $a$, $b$ two-by-two nonzero, nilpotent matrices. $ab$ must also be nilpotent, because of \eqref{idealGenP1}. It can be either zero or not.
\begin{description}
\item If $ab=0$, then $\im b\subseteq \ker a$, that again means $\im b\cong \ker a$. But $\im b \cong \ker b$ and $\im a \cong \ker a$. So, $a$ and $b$ have the same image and kernel. Hence $a \propto b$.
\item If $ab\neq 0$, then $\dim \im (ab)= \dim \ker (ab) = 1$. Moreover $\ker (ab)\supseteq \ker b$, that means $\ker (ab)\cong \ker b$, and $\im (ab)\subseteq \im a$, that means $\im (ab)\cong \im a$. This implies that $a$ and $b$ have the same image and kernel and then are proportional to each other.
\end{description} 
This proves that we can always bring $a,b,c,d$ in the form \eqref{nilp1} (in the phase $\vec{\theta}=\vec{\theta}_+$) or \eqref{nilp3} (in the phase $\vec{\theta}=\vec{\theta}_-$). Just as in Laufer's case, we could now set up an overarching Ansatz interpolating between the two phases, and allowing us to follow the simple length-two flop continuously.



\bibliography{mpa}
\bibliographystyle{at}

\end{document}